\documentclass[superscriptaddress]{revtex4-2}
\usepackage{graphicx}
\usepackage{subfig}
\usepackage{color}
\usepackage{amsmath}
\usepackage{amssymb}
\usepackage{bm}
\usepackage{slashed}
\usepackage{epsfig}
\usepackage{amsfonts}
\usepackage{epstopdf}
\usepackage{textcomp}
\usepackage{enumitem}
\usepackage{subfig}
\usepackage{hyperref}
\usepackage{cancel}
\usepackage{booktabs}
\usepackage{appendix}
\usepackage{amsthm}
\usepackage{mathrsfs}
\usepackage[table]{xcolor}
\usepackage{colortbl}
\usepackage{caption}

\setlist[itemize]{noitemsep, topsep=0pt} 

\begin{document}

\title{Enhancing the machine vision performance with multi-spectral light sources}

\author{Feng Zhang$^{a}$}
\email{zhangfeng@hes0501.com.cn}
\affiliation{
	Institute for Electric Light Sources, School of Information Science and Technology, Fudan University, Engineering Research Center of Advanced Lighting Technology, Ministry of Education, Shanghai 200433, China}
\affiliation{
	HES Technology Group Co., Ltd,
	Noble Center, 128 South 4th Ring West Road, Fengtai District, Beijing 100070, China}

\author{Rui Bao$^{a}$}
\email{baorui@hes0501.com.cn}
\affiliation{
	Institute for Electric Light Sources, School of Information Science and Technology, Fudan University, Engineering Research Center of Advanced Lighting Technology, Ministry of Education, Shanghai 200433, China}
\affiliation{
	HES Technology Group Co., Ltd,
	Noble Center, 128 South 4th Ring West Road, Fengtai District, Beijing 100070, China}

\author{Congqi Dai}
\email{daicongqi@hes0501.com.cn}
\affiliation{
	HES Technology Group Co., Ltd,
	Noble Center, 128 South 4th Ring West Road, Fengtai District, Beijing 100070, China}

\author{Wanlu Zhang}
\email{fdwlzhang@fudan.edu.cn}
\affiliation{
	Institute for Electric Light Sources, School of Information Science and Technology, Fudan University, Engineering Research Center of Advanced Lighting Technology, Ministry of Education, Shanghai 200433, China}

\author{Shu Liu$^{b}$}
\email{liushu@hes0501.com.cn}
\affiliation{
	HES Technology Group Co., Ltd,
	Noble Center, 128 South 4th Ring West Road, Fengtai District, Beijing 100070, China}

\author{Ruiqian Guo$^{b}$}
\email{rqguo@fudan.edu.cn}
\thanks{\\ a: Equivalent contribution authors; b: Corresponding authors. }
\affiliation{
	Institute for Electric Light Sources, School of Information Science and Technology, Fudan University, Engineering Research Center of Advanced Lighting Technology, Ministry of Education, Shanghai 200433, China}

\begin{abstract}
This study mainly focuses on the performance of different multi-spectral light sources on different object colors in machine vision and tries to enhance machine vision with multi-spectral light sources. 
Using different color pencils as samples, by recognizing the collected images with two classical neural networks, AlexNet and VGG19, the performance was investigated under 35 different multi-spectral light sources. The results show that for both models there are always some non-pure white light sources, whose accuracy is better than pure white light, which suggests the potential of multi-spectral light sources to further enhance the effectiveness of machine vision. The comparison of both models is also performed, and surprised to find that the overall performance of VGG19 is lower than that of AlexNet, which shows that the importance of the choice of multi-spectral light sources and models. 

\end{abstract}

\maketitle
\allowdisplaybreaks

\section{Introduction}
Machine vision, in general, refers to the use of machines instead of the human eye to perform a range of visual tasks, which mainly includes image collection and processing technology. The former requires the use of a combination of light sources and acquisition devices, while the latter requires well-designed and trained algorithms.

In the last decade, image processing technology has developed rapidly, and Artificial Intelligence algorithms, represented by deep learning, have become the mainstream of image processing methods. deep learning is the process of learning image features by training a multi-layer neural network model using large amounts of data. deep learning in machine vision began with ALexNet~\cite{WOS:000402555400026}, and then a variety of more sophisticated network architectures were devised, such as VGG~\cite{simonyan2015deep}, ResNet~\cite{WOS:000400012300083}, etc. Nowadays, deep learning performs very well in various image tasks and achieves significant results~\cite{WOS:000868715700001}. 

The algorithm is one aspect of influencing the machine vision effect, while the image quality is more important. It has been reported that image contrast can significantly affect machine vision results, and the accuracy of all networks decreases in low-contrast images~\cite{akbarinia2018contrast,akbarinia2019manifestation}. Another study considered the impact of five different types of image quality distortion on deep neural networks and showed that the distortion of the images had a significant impact on the performance of all networks, with the more severe the distortion, the greater the impact~\cite{WOS:000391251500036}. 

Many studies are comparing the similarities and differences between human vision and machine vision~\cite{WOS:000461852002012,WOS:000463806000105,NEURIPS2020_9f699296,WOS:000922928402048,wichmann2017methods}, and the results show that machine vision is comparable to human vision in the case of clear images, but much less robust in the case of distorted images. The differences could indicate that the internal mechanisms of image processing are different between neural networks and human visual system.

The light source is an indispensable part of object imaging, except for a few luminous objects, objects form vision by reflecting light sources. Therefore, whether for machine vision or human vision, the quality of the light source will greatly affect the quality of imaging. A review of the light source and illumination system design can be found in~\cite{WOS:000654942100001}.

The human eye can only see visible light, and the color of light is one of the important properties of light. Physiologically speaking, the color of light is the response formed in the brain when light enters the human eye through analysis and processing; physically speaking, different colors of light reflect different spectral shapes. In general, white light is considered the most beneficial to human vision for which has the best color rendering. However, as the different performances in terms of robustness, the optimal color of light may also not be right for machine vision, or rather, the correctness needs to be tested. 

As mentioned in~\cite{WOS:000654942100001}, in some scenes, specific colors of light may make machine vision perform better. A review of light source color selection and optimization methods can be found in~\cite{WOS:000387429500117}. However, these selection methods are from the viewpoint of grayscale images, which is mainly suitable for traditional machine vision algorithms. The deep learning algorithms are directly processing RGB images, which have much more information. In addition, the color properties of the object itself, which is a concentrated expression of the spectral properties of the object, are not fully applied, and the relationship between the object color and the color of the optimal light source for machine vision is not confirmed. 

Therefore, this study mainly focuses on the performance of multi-spectral light sources on different object colors in machine vision. And the paper is organized as follows: In Section~\ref{sec:method}, we introduce our research materials and methods. And then in Section~\ref{sec:result}, we give our experimental and analytical results. In response to the results, the discussion is given in Section~\ref{sec:discussion}. The summary is given in the final section.  

\section{Methodology}\label{sec:method}
\subsection{Machine Vision Algorithms}
The machine vision algorithms here are deep learning algorithms. Because of the complex, opaque, and black-box nature of the deep neural networks~\cite{WOS:000642759800001,WOS:000427742400005}, we take an experimental approach to test the effect of different spectra of machine vision in this study. We studied two distinguished deep learning image classification models: AlexNet~\cite{WOS:000402555400026,krizhevsky2014weird} and VGG19\_BN~\cite{simonyan2015deep}, named VGG19 in the following. The weights were obtained from PyTorch platform~\cite{WOS:000534424308009} and set to 'DEFAULT'. It is worth emphasizing that the weights of all neural networks are trained on the ImageNet-1K dataset~\cite{WOS:000365089800001}, which means the networks can predict 1000 labels and the object to be predicted should preferably be in the 1000 classes. 

The reason for choosing these two models: 
\begin{itemize}
	\item  Image classification tasks are the basis for other vision tasks such as object detection and image segmentation, and classification models are also the basis for other vision models. Thus the structure of the classification model is widely distributed over other deep learning models. 
	\item  They are classical models with a typical structure that will lead to general conclusions.  
	\item  These two models have different structures and weight scales, and the comparison of the results may lead to new findings.
\end{itemize}

\subsection{Multi-spectral Light Source}
\begin{figure}[ht]
	\centering
	\subfloat[Red Channel]{
		\includegraphics[width=5.6cm]{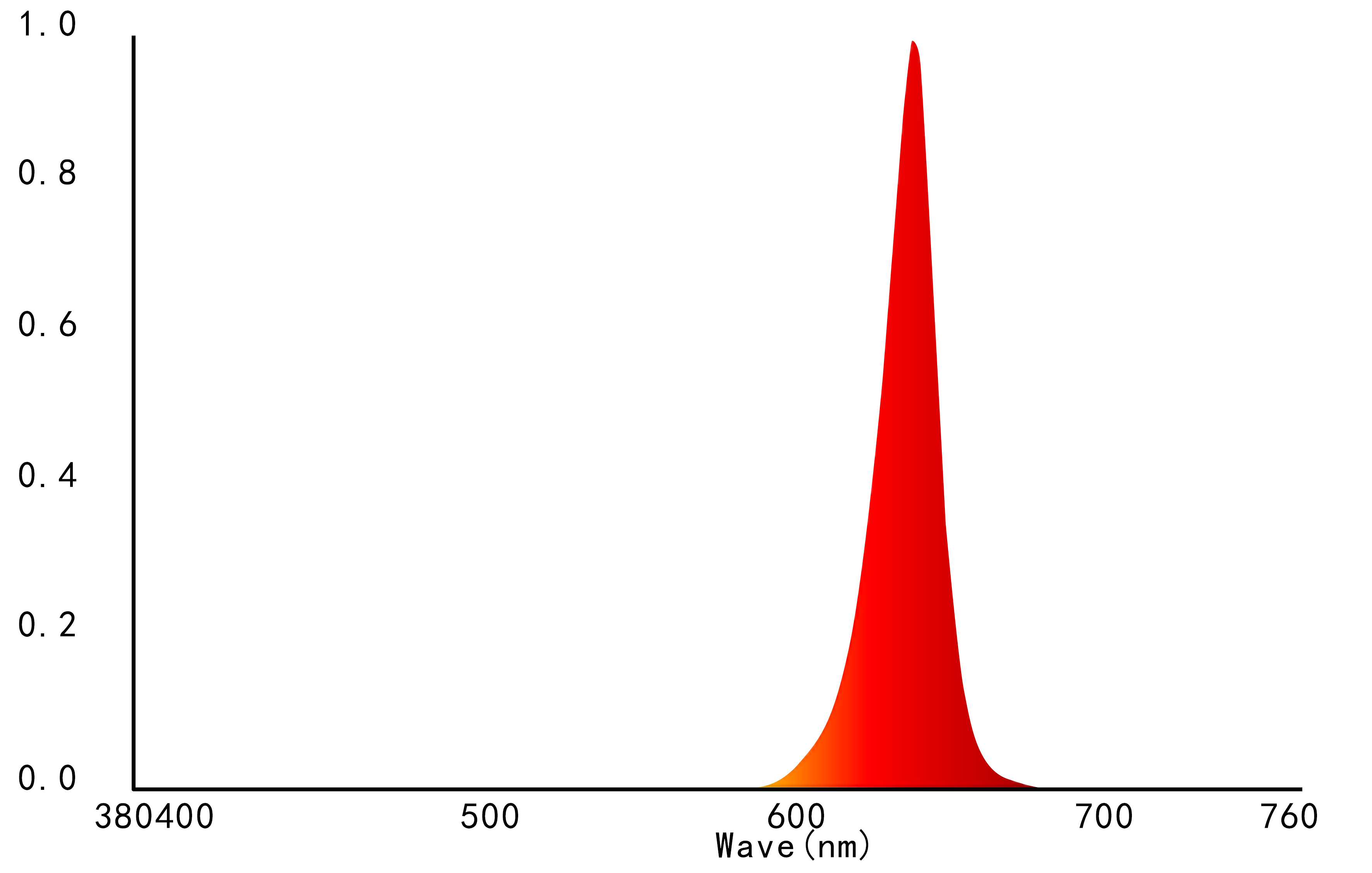}
	}
	\subfloat[Green Channel]{
		\includegraphics[width=5.6cm]{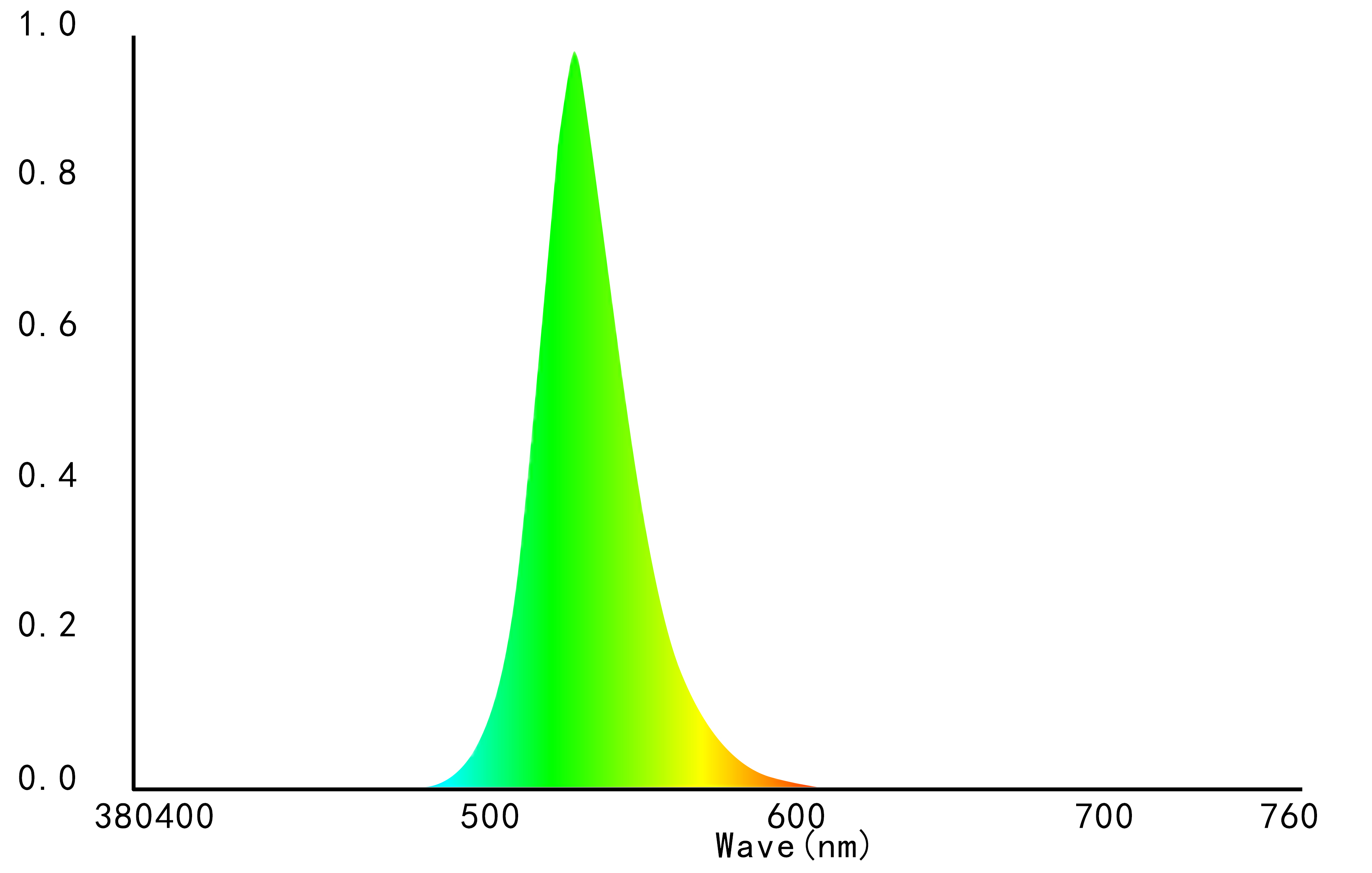}
	}
	\subfloat[Blue Channel]{
		\includegraphics[width=5.6cm]{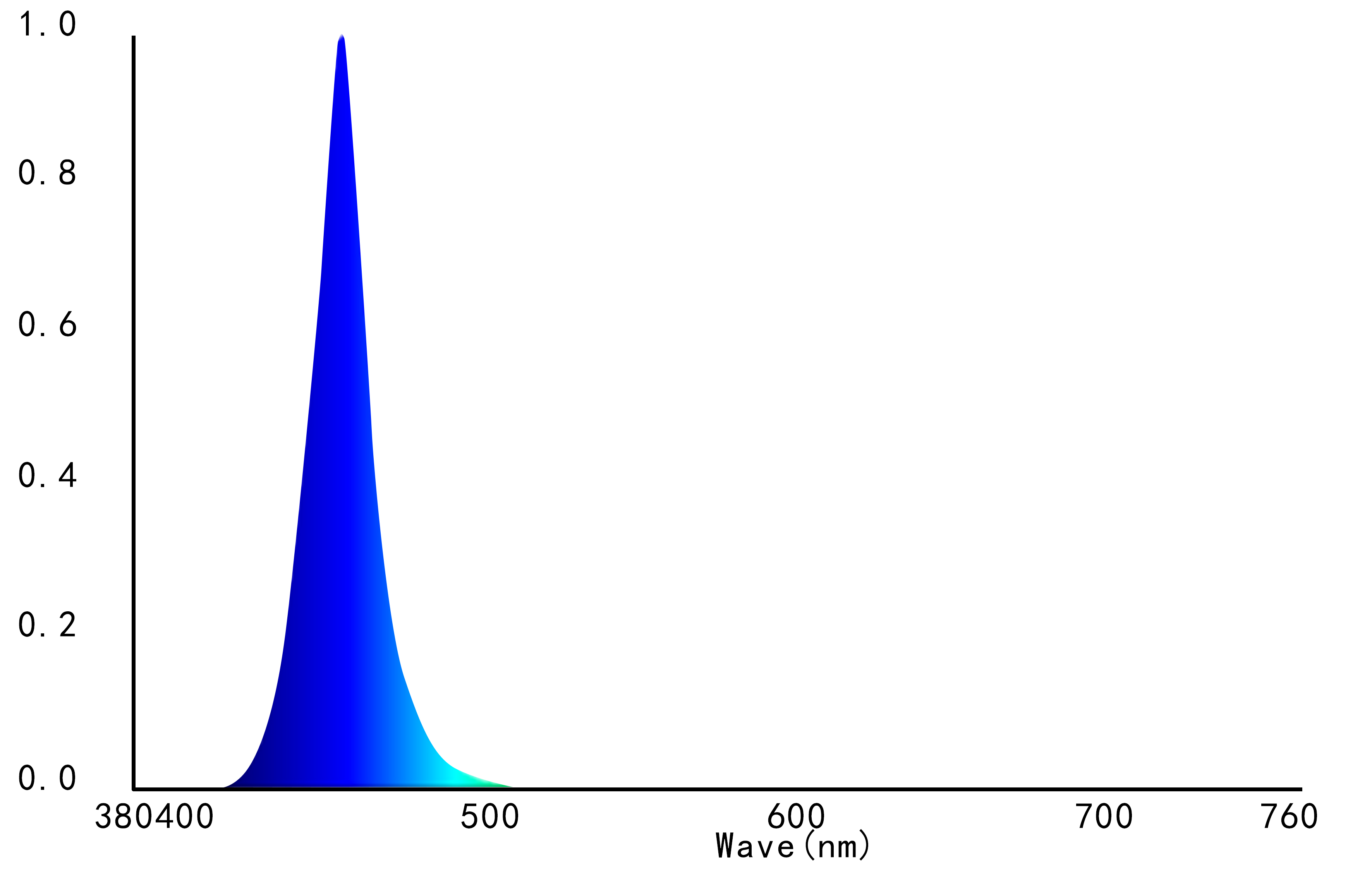}
	}\\
	\subfloat[Amber Channel]{
		\includegraphics[width=5.6cm]{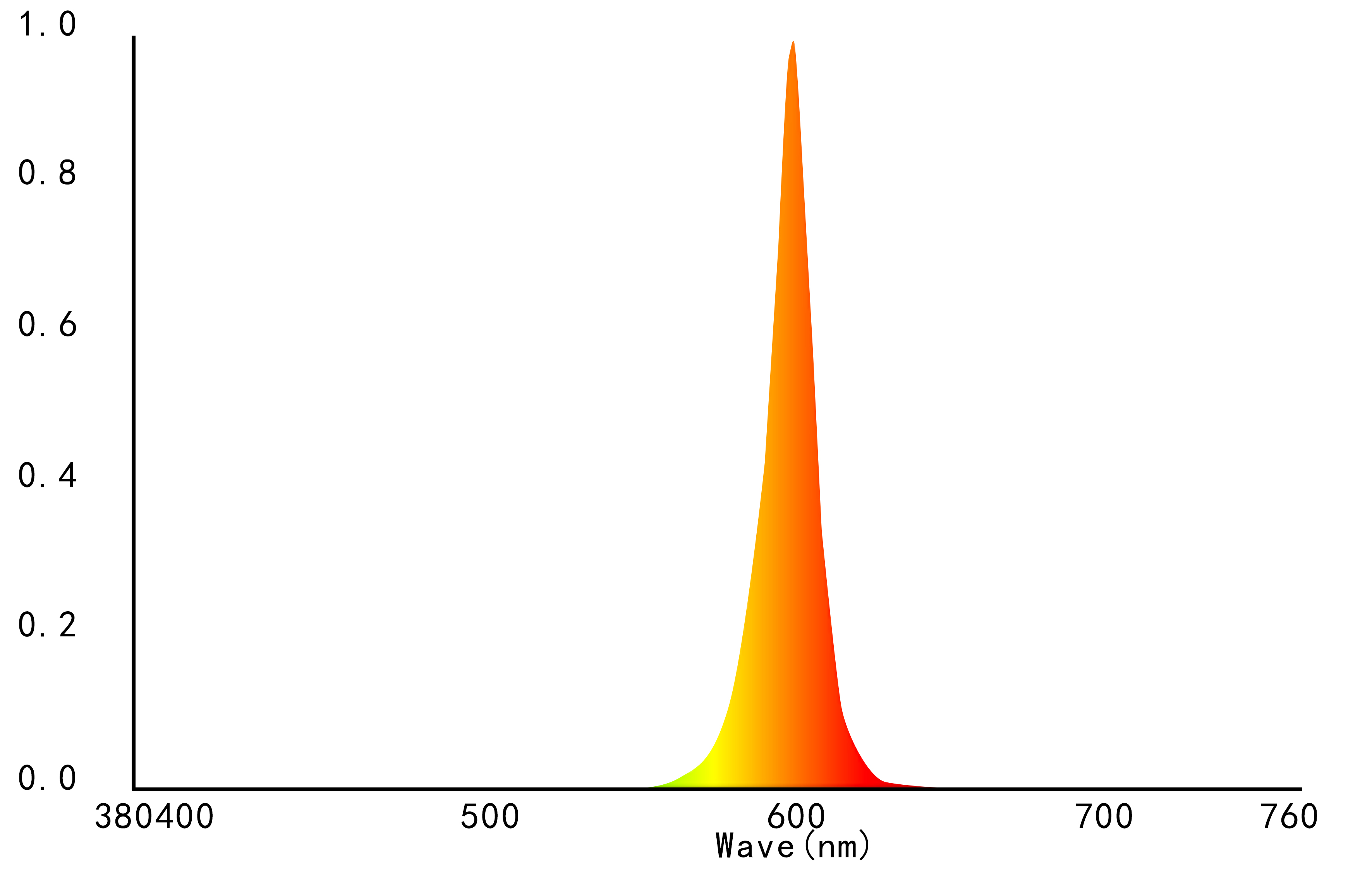}
	}
	\subfloat[Cyan Channel]{
		\includegraphics[width=5.6cm]{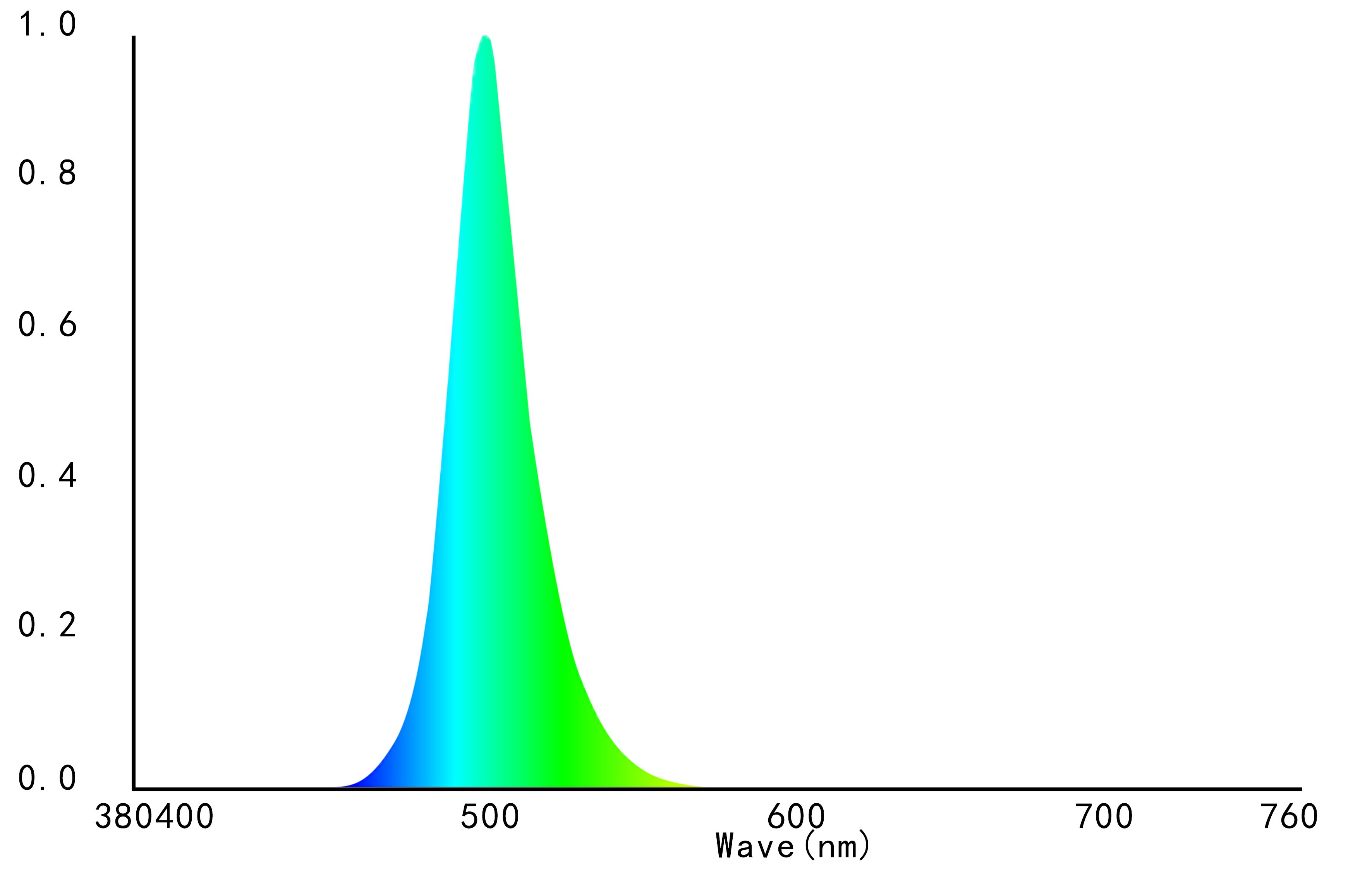}
	}
	\caption{The spectral distribution of each channel of the customized five-channel lamp.}
	\label{sp_dis}
\end{figure}

The multi-spectral light source is generated by a customized five-channel lamp with the spectral distribution of each channel as shown in FIG.~\ref{sp_dis}. The output ratio of each channel can be controlled to produce different multi-spectral or color light sources, and the control is performed through the Processing program following the DMX512 lighting control protocol. 

\begin{figure}[ht]
	\centering
	\includegraphics[width=9cm]{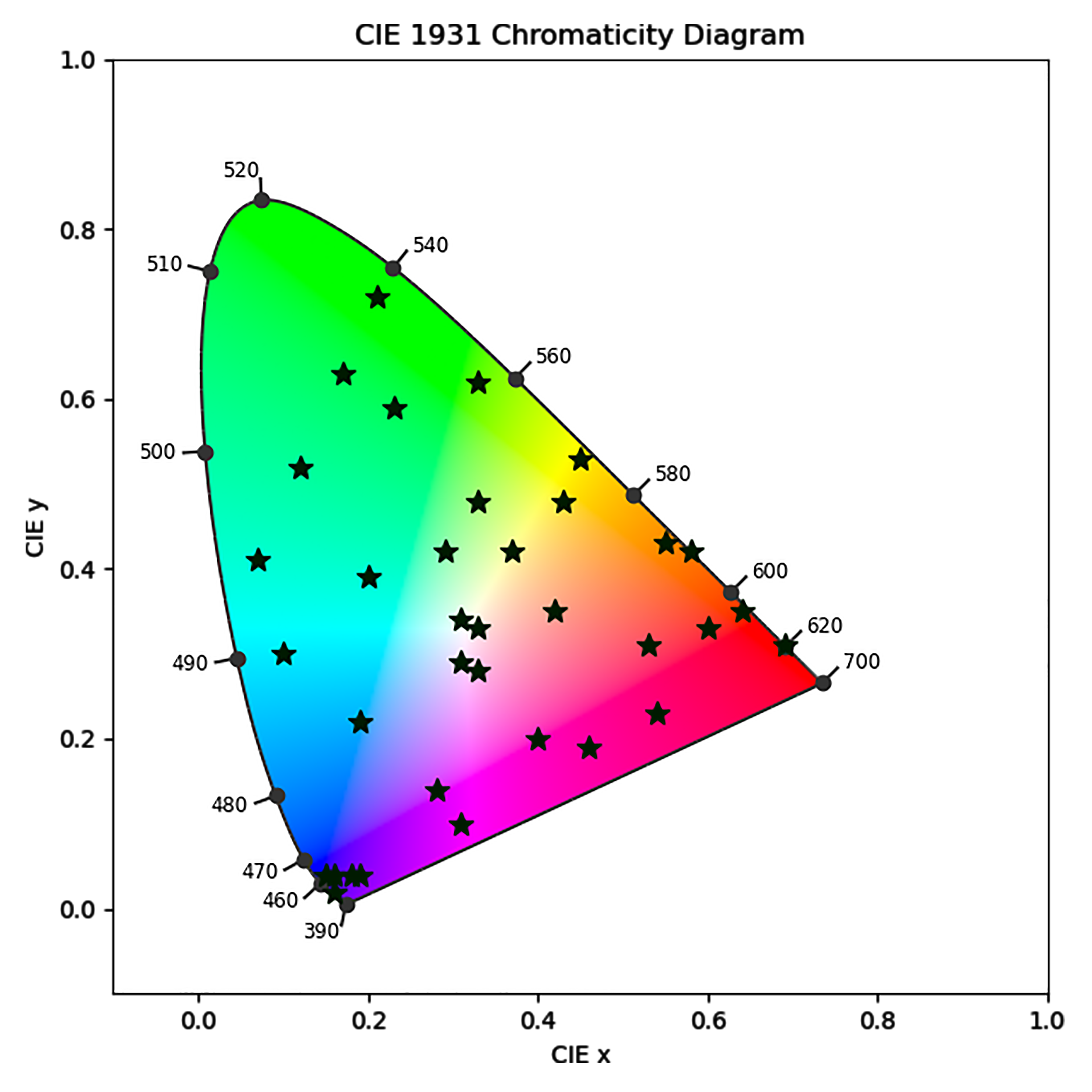}
	\caption{CIE 1931 Chromaticity Diagram and 35 multi-spectral light sources used in the experiment.}
	\label{cie1931_allpoint}
\end{figure}

The description of the different spectral colors can be done with the CIE standard colorimetric system, of which the most classic and commonly used is the CIE 1931 standard colorimetric system~\cite{smith1931cie}. CIE 1931 Chromaticity Diagram can be seen in FIG.~\ref{cie1931_allpoint}, where each multi-spectral light source has a unique color coordinate corresponding to it, and the color coordinates of the 35 multi-spectral light sources used in the experiment are also marked with black pentagrams. The illuminance of these multi-spectral light sources is kept the same, differing only in color coordinates. In the experiment, the illuminance value of the light source to the object surface is set to 30 lx.
\subsection{Experimental Objects of Different Colors}
The study mainly focuses on the performance of multi-spectral light sources on different object colors in machine vision. The following considerations were made regarding the selection of experimental objects:
\begin{itemize}
	\item  The objects need to be of different colors and not different from each other except for the color. 
	\item  In order to obtain quantitative data for different multi-spectral contrasts, a certain number of images need to be acquired for each object color and each multi-spectral light source.  
	\item  At the same time, in order to control a single variable, the difference between acquired images should exist only in the color of the object and the multi-spectral light source, which means that all positions of the image acquisition device, including the position of the object, the light source and the camera, should be kept constant. Therefore, the problem of the number of images must be solved by other means.
\end{itemize}

\begin{figure}[ht]
	\centering
	\includegraphics[width=9cm]{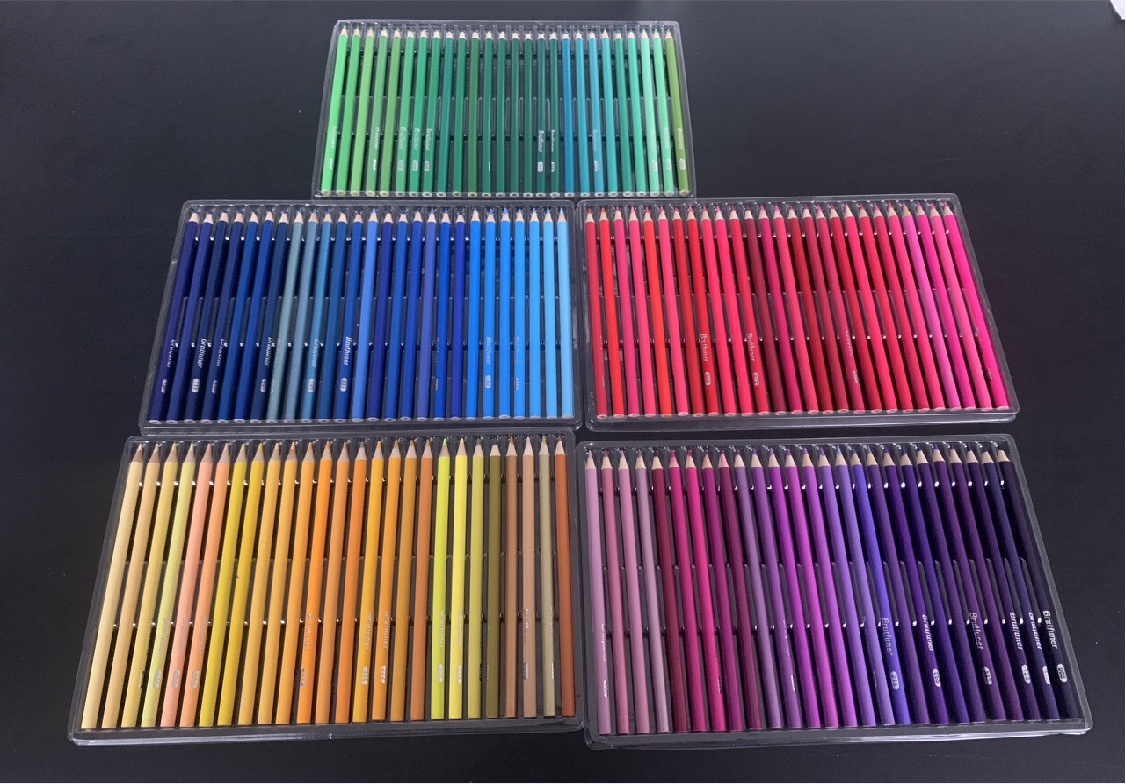}
	\caption{Experimental objects: different color pencils.}
	\label{pencil}
\end{figure}

Based on the above considerations, different color pencils are a very good choice, as shown in FIG.~\ref{pencil}. According to the Munsell Color System~\cite{cochrane2014munsell}, the different color pencils are divided into five sub-classes, where each sub-class represents a Munsell primary hue. There are five primary hues: green, blue, red, green, and purple. For each primary hue, there are 26 pencils, all of which belong to the primary hue but with slight differences in color. The 26 pencils in each primary hue are considered as a whole, as the source of the number of samples in this primary hue.  

\subsection{Collection of Image Datasets}
\begin{figure}[ht]
	\centering
	\subfloat{
		\includegraphics[width=9cm]{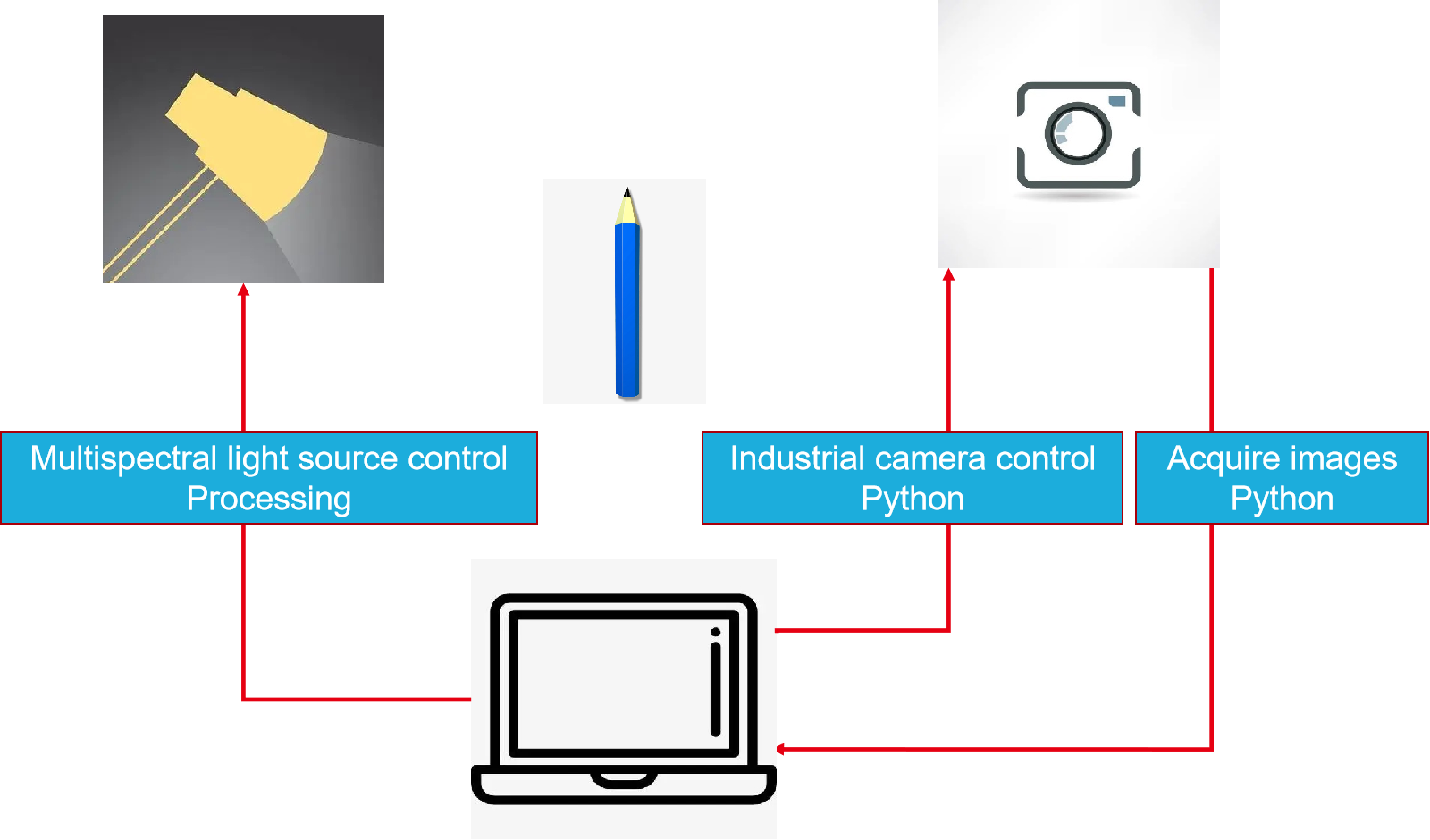}
	}
	\subfloat{
		\includegraphics[width=7.5cm]{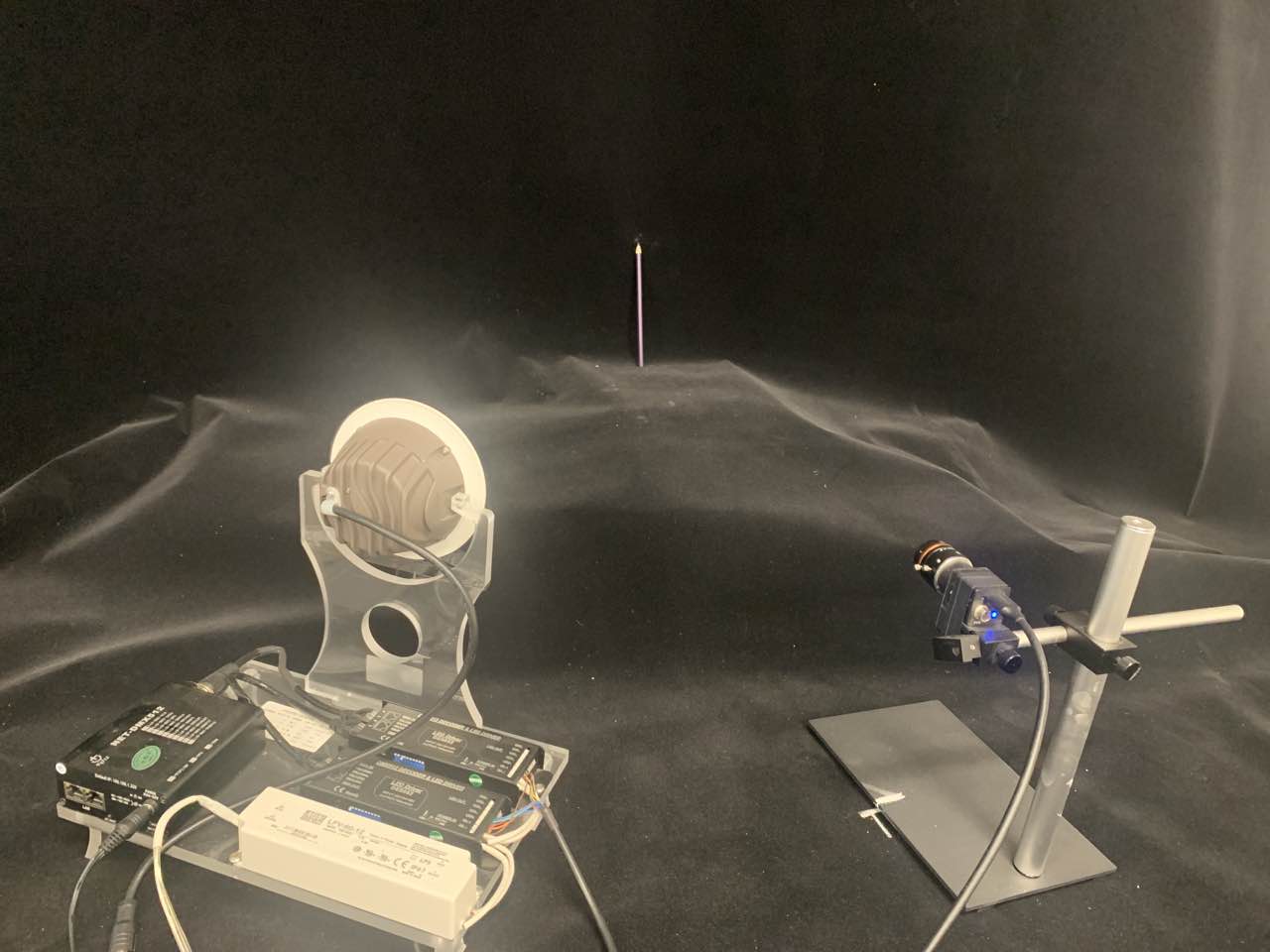}
	}
	\caption{The schematic and realistic picture of the image collection system.}
	\label{collection}
\end{figure}

The image collection system consists of a computer-controlled program, a multi-spectral light source, an industrial camera and experimental objects, whose schematic and realistic picture are shown in FIG.~\ref{collection}. The camera used in the experiment is a HIKVISION industrial camera, and the product model number is MV-CE060-10UC, which has a resolution of 3072 $\times$ 2048. The distance from the camera and the light source to the object is 70 cm, with the camera facing the object and the light source at 30° to the camera.

\begin{figure}[ht]
	\centering
	\subfloat[Red Channel]{
		\includegraphics[width=5.6cm]{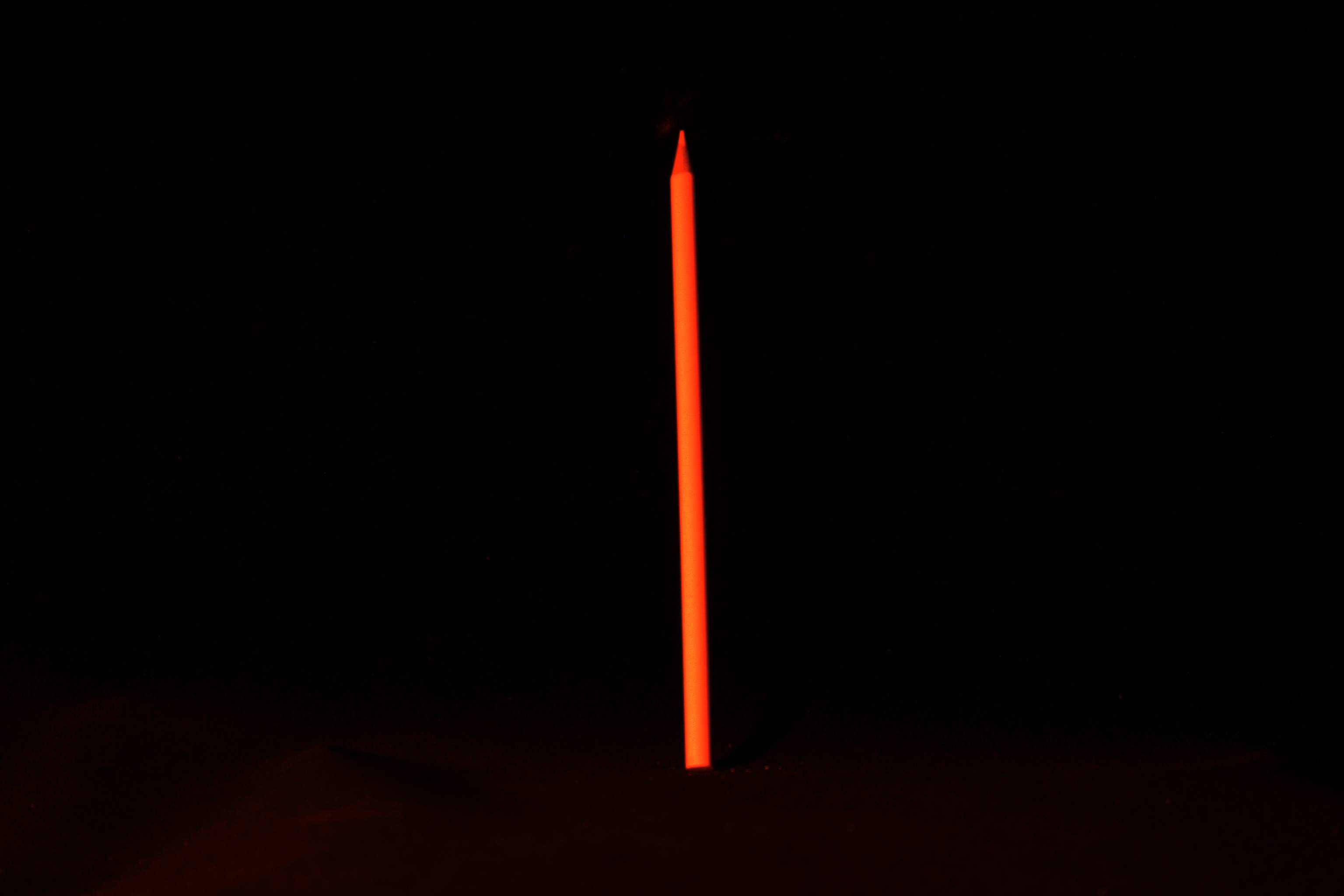}
	}
	\subfloat[Green Channel]{
		\includegraphics[width=5.6cm]{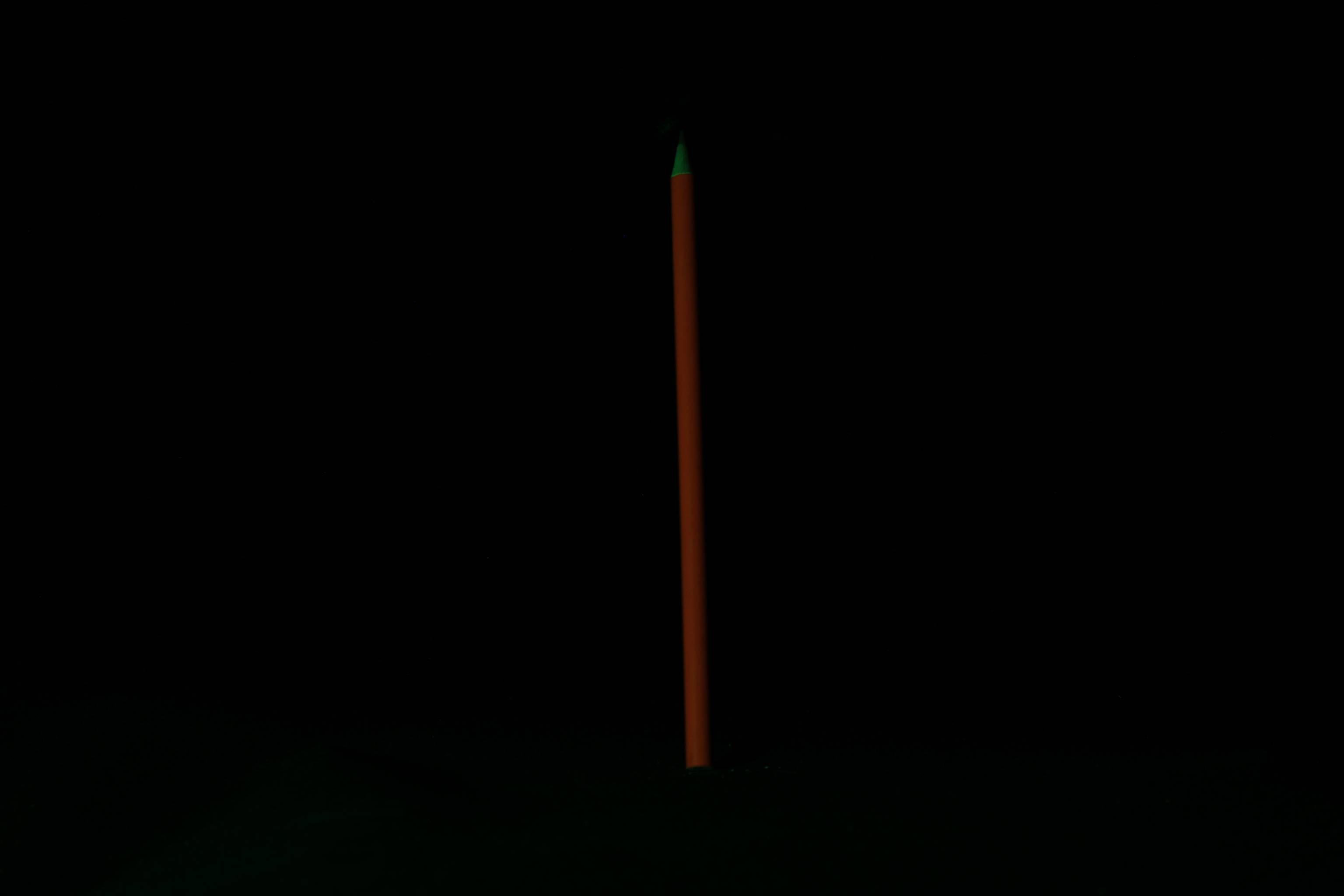}
	}
	\subfloat[Blue Channel]{
		\includegraphics[width=5.6cm]{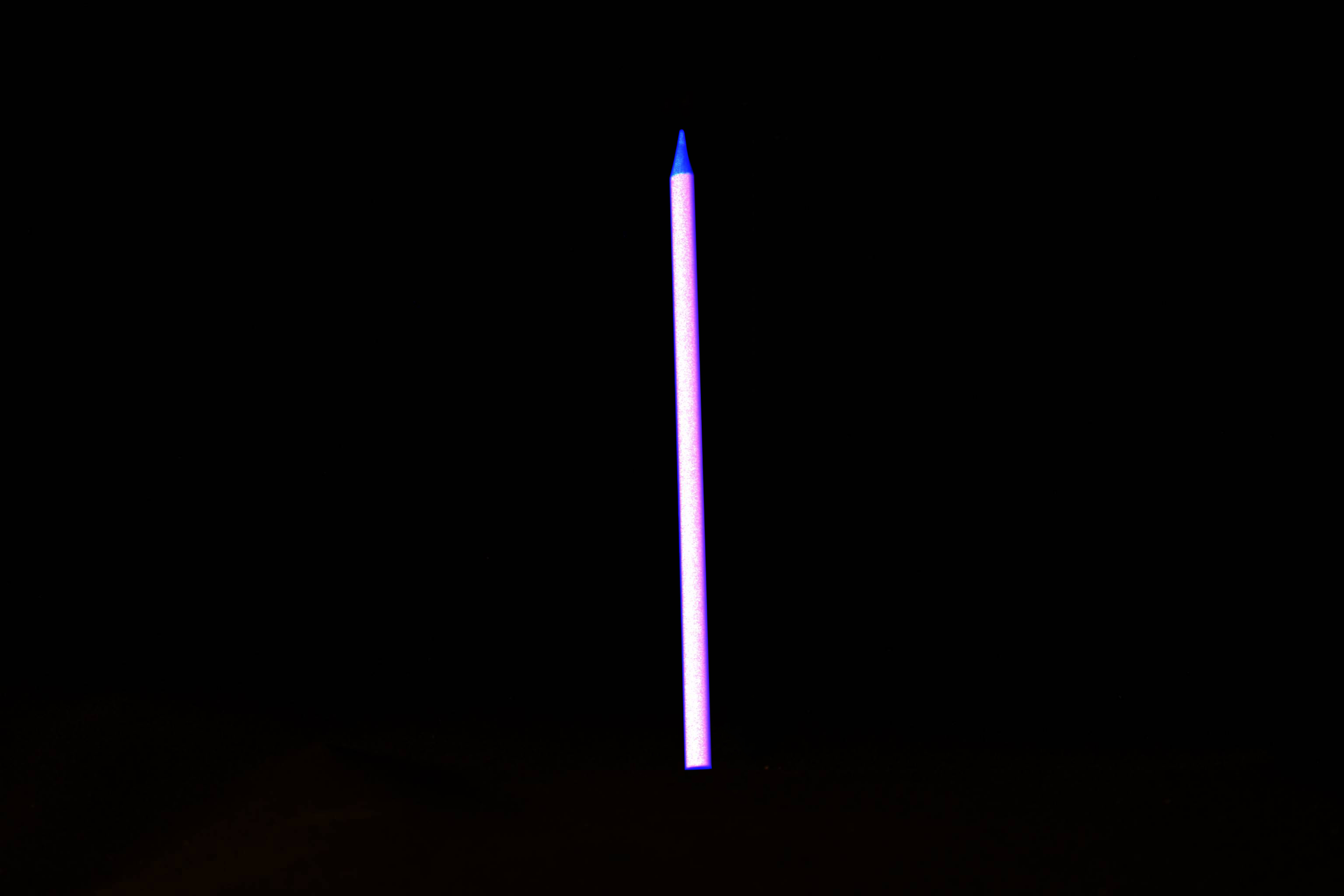}
	}\\
	\subfloat[Amber Channel]{
		\includegraphics[width=5.6cm]{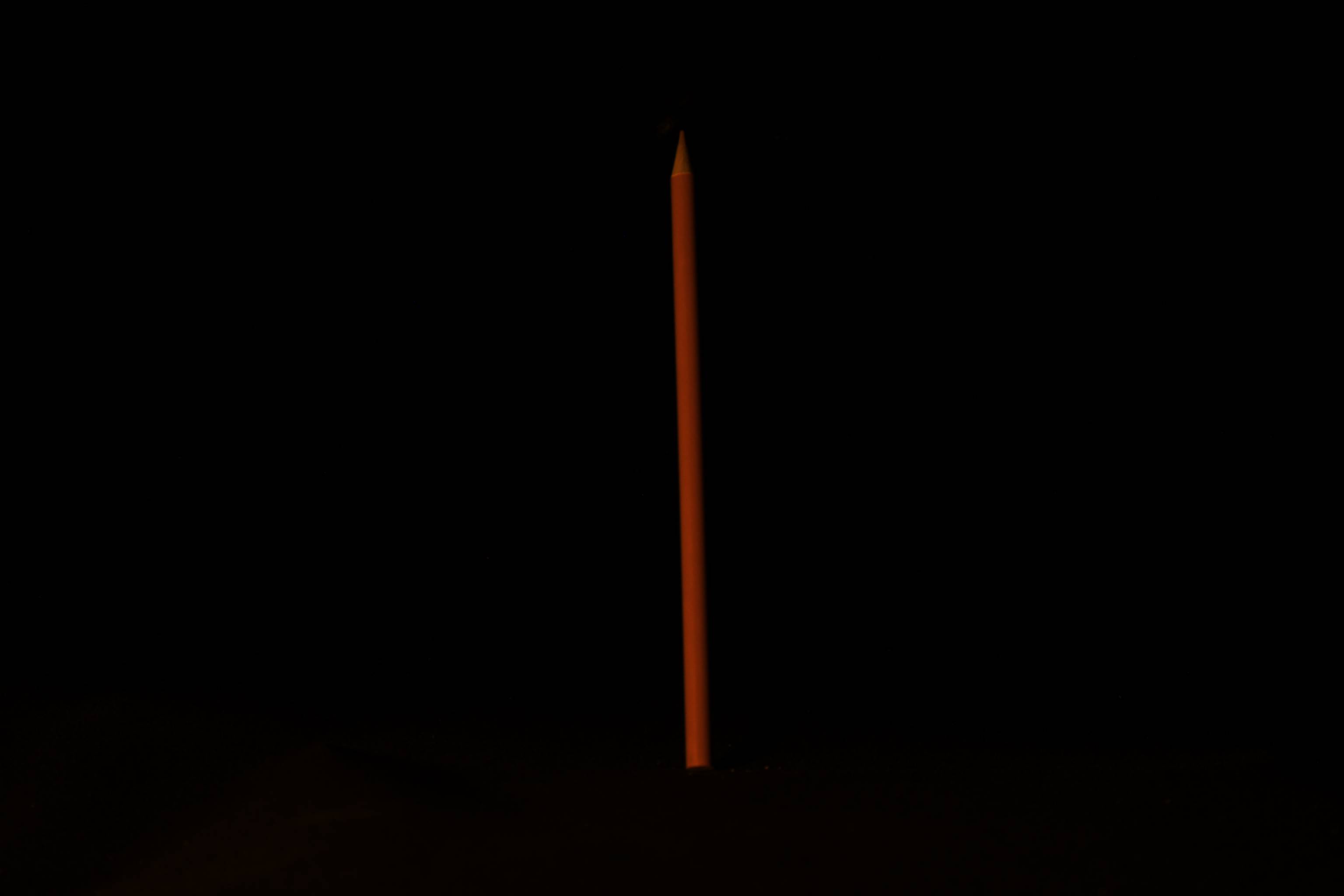}
	}
	\subfloat[Cyan Channel]{
		\includegraphics[width=5.6cm]{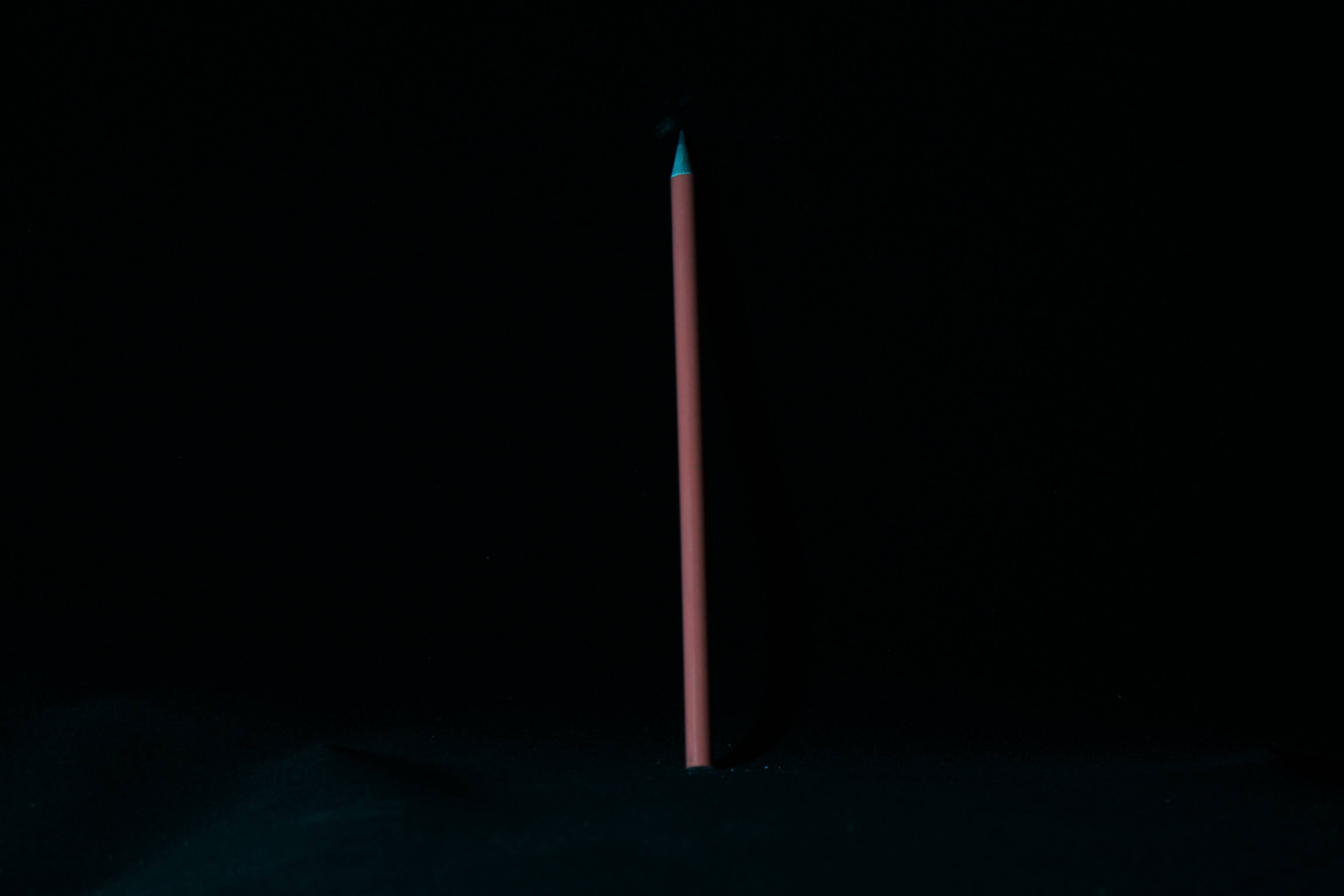}
	}
	\caption{Example of the collected images: a red pencil under five single-channel multi-spectral lights as shown in FIG.~\ref{sp_dis}. }
	\label{example}
\end{figure}

The image collection system follows the following process:
\begin{itemize}
	\item step 1: Fix objects, for a specific multi-spectral light source, the Processing program is used to control the multi-spectral light source, and the Python program is used to control the industrial camera to complete the image acquisition and information recording.
	\item step 2: Change the light source to complete the collection of images corresponding to all 35 multi-spectral light sources, and the number of images is 35. 
	\item step 3: Change the object and repeat the previous steps to complete the collection of all images.
\end{itemize}
Finally, the collected images can be formed into an image table as TABLE~\ref{table:img_tab}, where each cell represents an image, which is obtained in the corresponding multi-spectral light source for the corresponding color pencil, and five of them, as an example, can be seen in FIG.~\ref{example}. These five images were obtained by a red pencil under five single-channel multi-spectral lights as shown in FIG.~\ref{sp_dis}, and it can be seen that the same object is imaged very differently in different multi-spectral light sources.

\begin{minipage}{\textwidth}
	\centering
	\includegraphics[width=17cm]{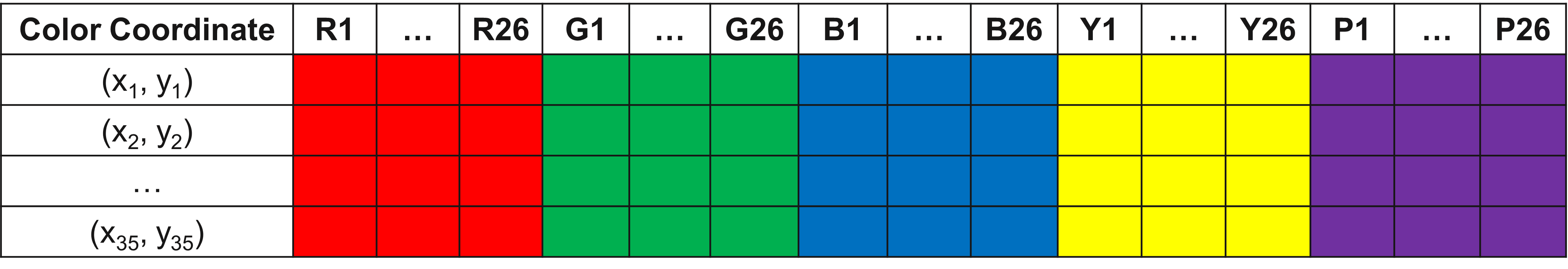}
	\captionof{table}{Table of the collected images.}
	\label{table:img_tab}
\end{minipage}

\subsection{Datasets Analysis}
For a particular image, a trained model can be used to predict its labels. Note that before the image is fed into the neural network, some pre-processing steps are required: 
\begin{itemize}
	\item First, the images are cropped to the center square according to the input size of each network, and for both AlexNet and VGG19 used in this study, the size is 224 $\times$ 224.
	\item Second, it is pre-processed in the same way as used in the original training process of this network.
\end{itemize}

For this study, the correct label should be 'pencil'. However, the pencil is not in the training set of ImageNet-1K, which means the label 'pencil' cannot be predicted. One straightforward approach is to use the dataset containing the pencil to rebuild and train the model, and another roundabout approach is to look for labels on the training set that are similar under human vision as the correct labels. The reason for the latter is that among a limited number of labels, the prediction should be considered correct as long as the labels predicted by the model are consistent with the labels selected by human vision. The latter roundabout approach was adopted in this study, and the direct approach was left for further studies.

Similar labels to 'pencil' in ImageNet-1K are 'ballpoint', 'fountain\_pen', 'matchstick', and 'pole'. If the predicted label is in the similar labels, the prediction for this image is considered to be correct. Judging all the images of TABLE~\ref{table:img_tab}, a similar table can be obtained, and each cell in the table is the judgment of the corresponding image: True or False. 

For a certain multi-spectral light source with color coordinate $(x_i, y_i)$, the accuracy of certain pencil color $C$ is
\begin{equation}
	Acc((x_i, y_i),C)=\dfrac{\text{Num of Correct}((x_i, y_i),C)}{26},
\end{equation}
where $\text{Num of Correct}((x_i, y_i),C)$ is the number of correct predictions for the pencil color $C$, under the multi-spectral light source with color coordinate $(x_i, y_i)$. The accuracy of all pencil colors is
\begin{equation}
	Acc((x_i, y_i))=\dfrac{\sum_{C}\text{Num of Correct}((x_i, y_i),C)}{26 \times 5},
\end{equation}
for all the primary hues are included. 
\section{Results}\label{sec:result}
Here we present the results of two classification models: AlexNet and VGG19. 
\subsection{AlexNet}
\begin{figure}[ht]
	\centering
	\subfloat[Red Pencil]{
		\includegraphics[width=5.2cm]{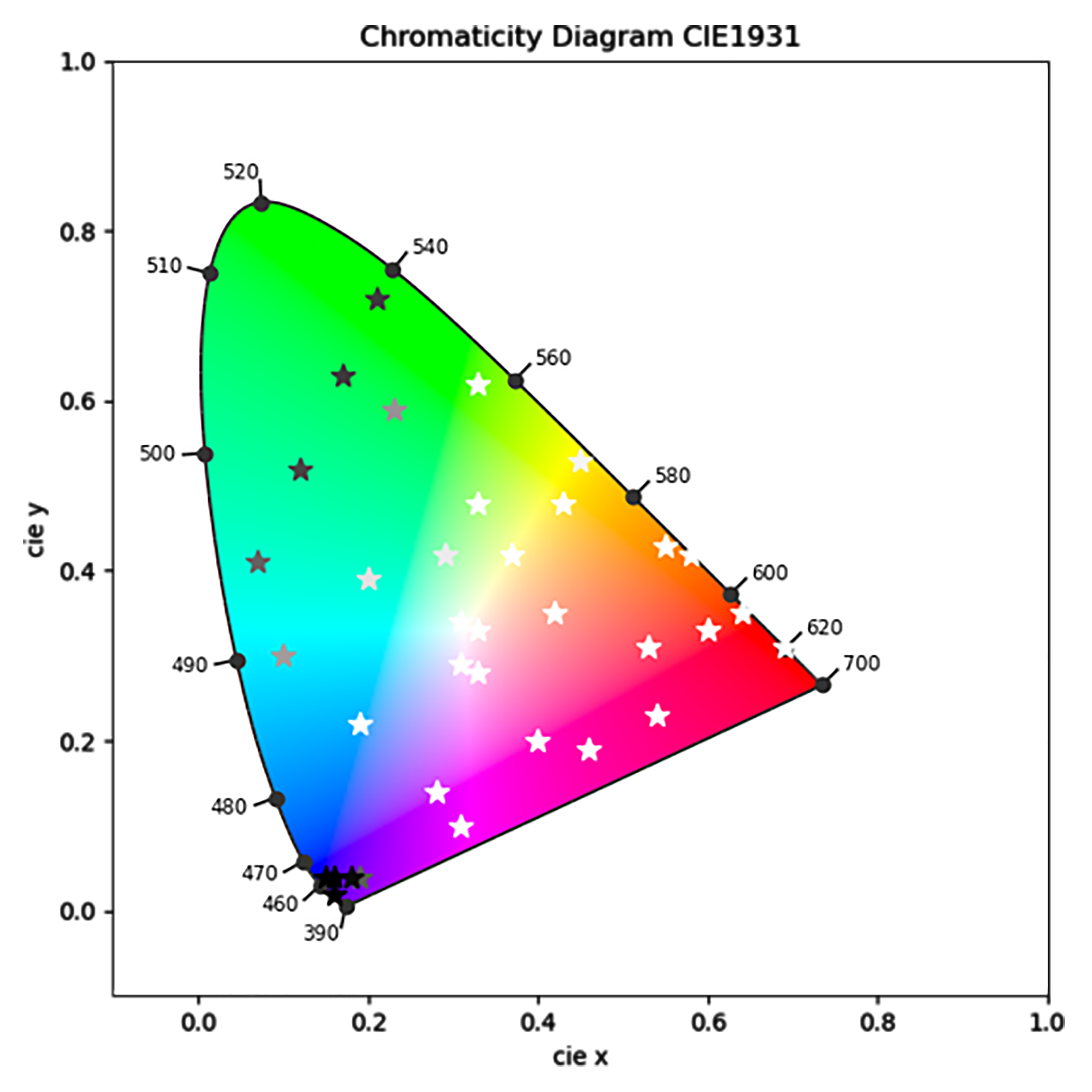}
	}
	\subfloat[Green Pencil]{
		\includegraphics[width=5.2cm]{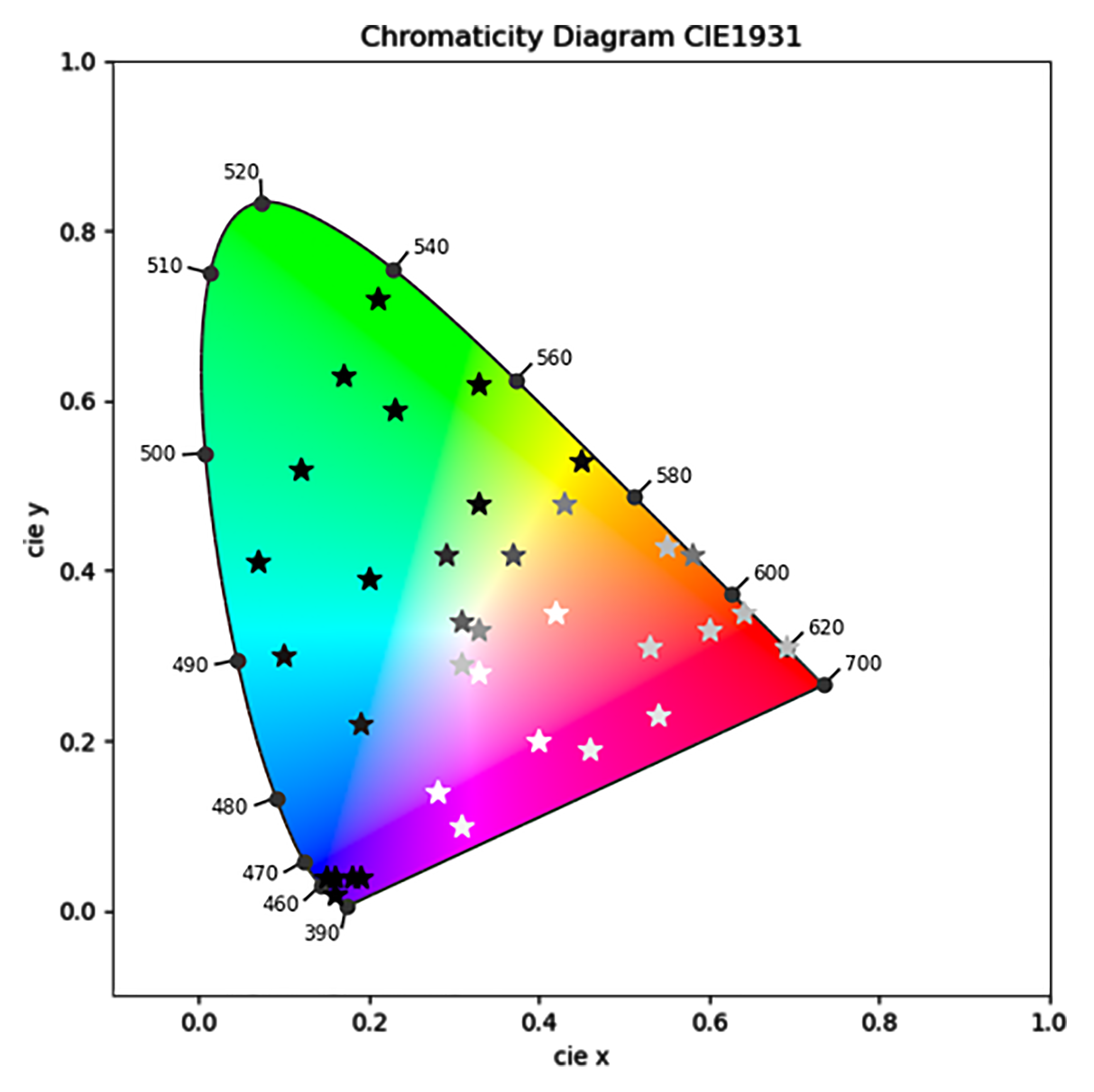}
	}
	\subfloat[Blue Pencil]{
		\includegraphics[width=5.2cm]{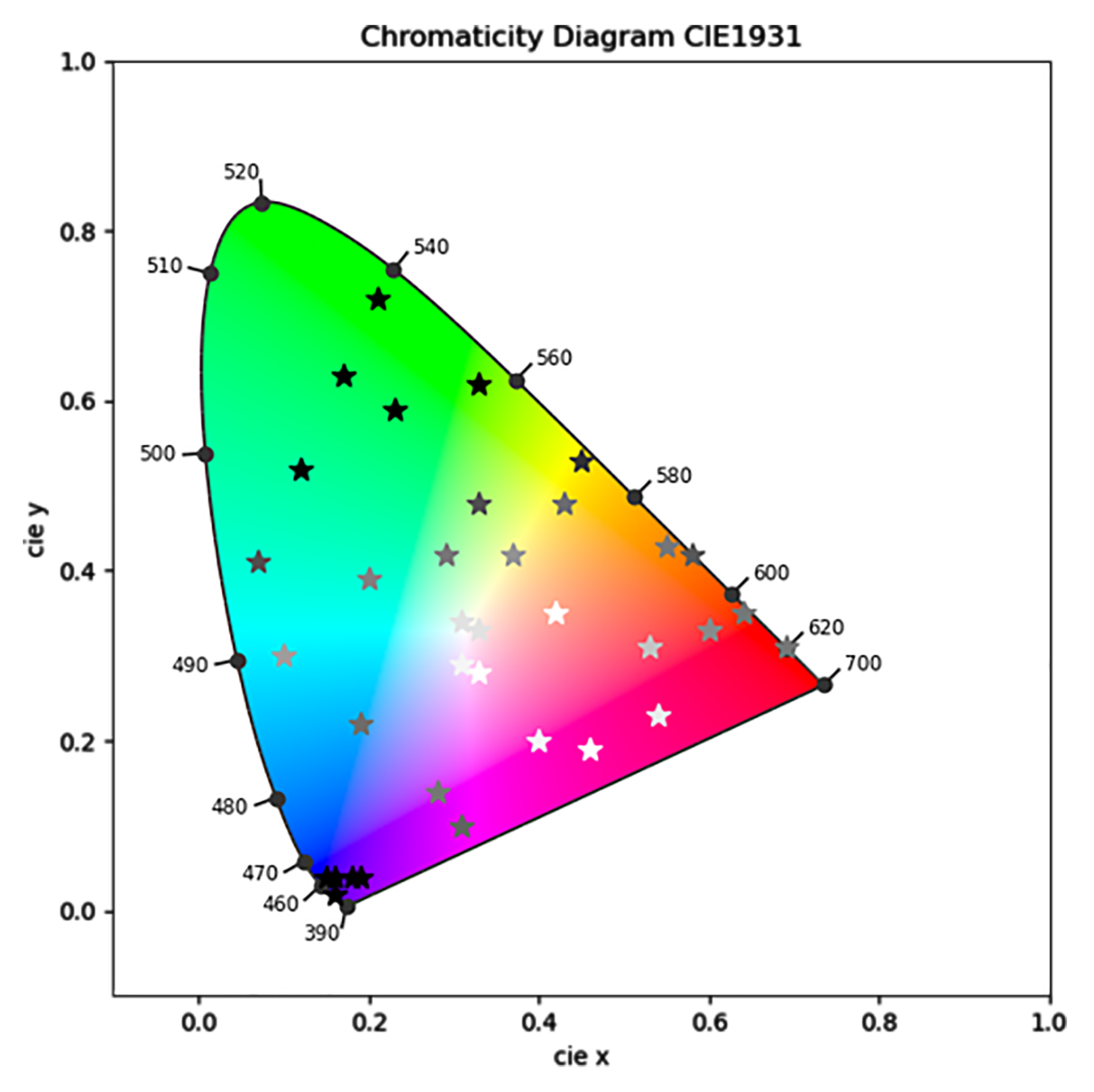}
	}\\
	\subfloat[Yellow Pencil]{
		\includegraphics[width=5.2cm]{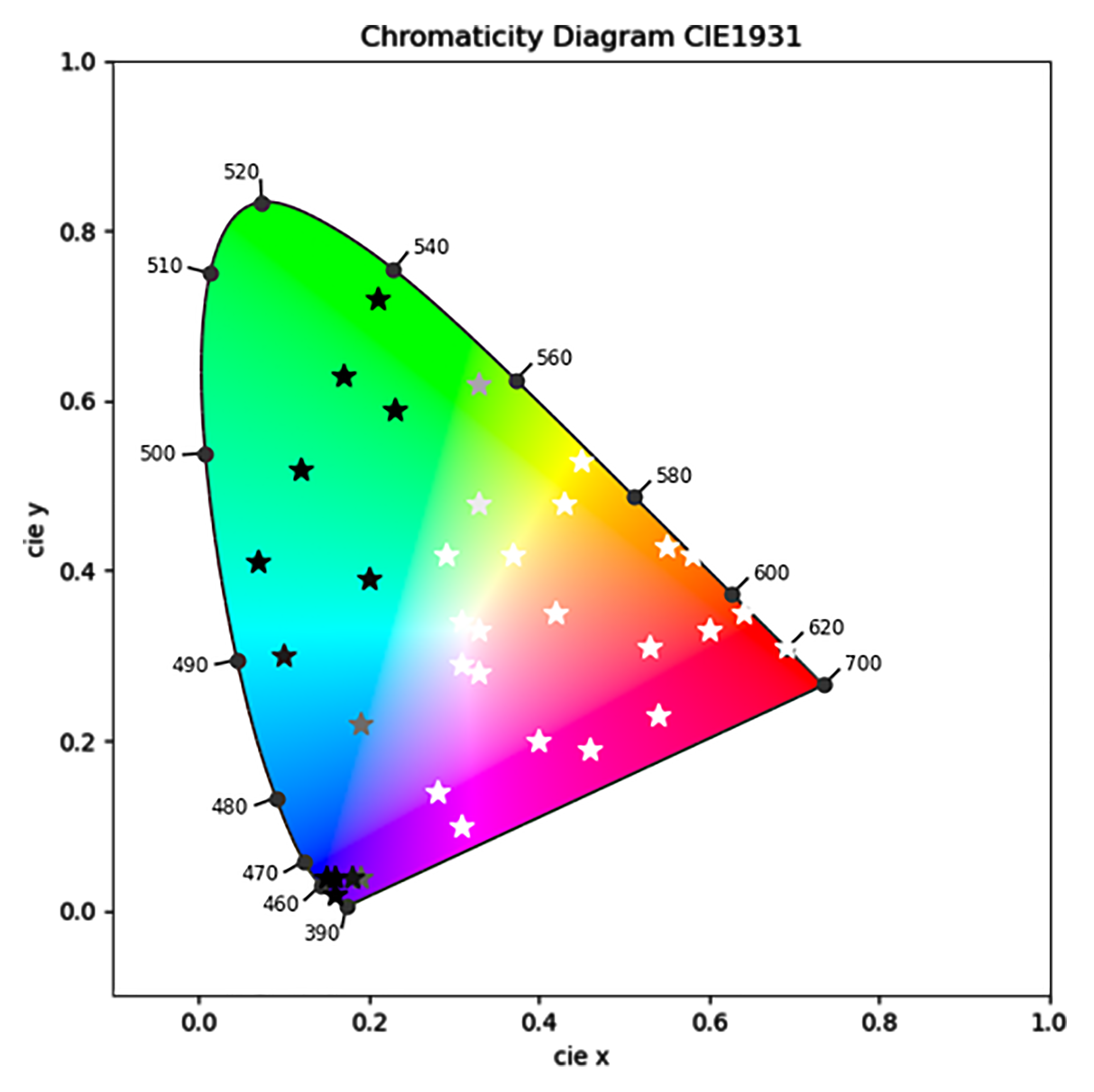}
	}
	\subfloat[Purple Pencil]{
		\includegraphics[width=5.2cm]{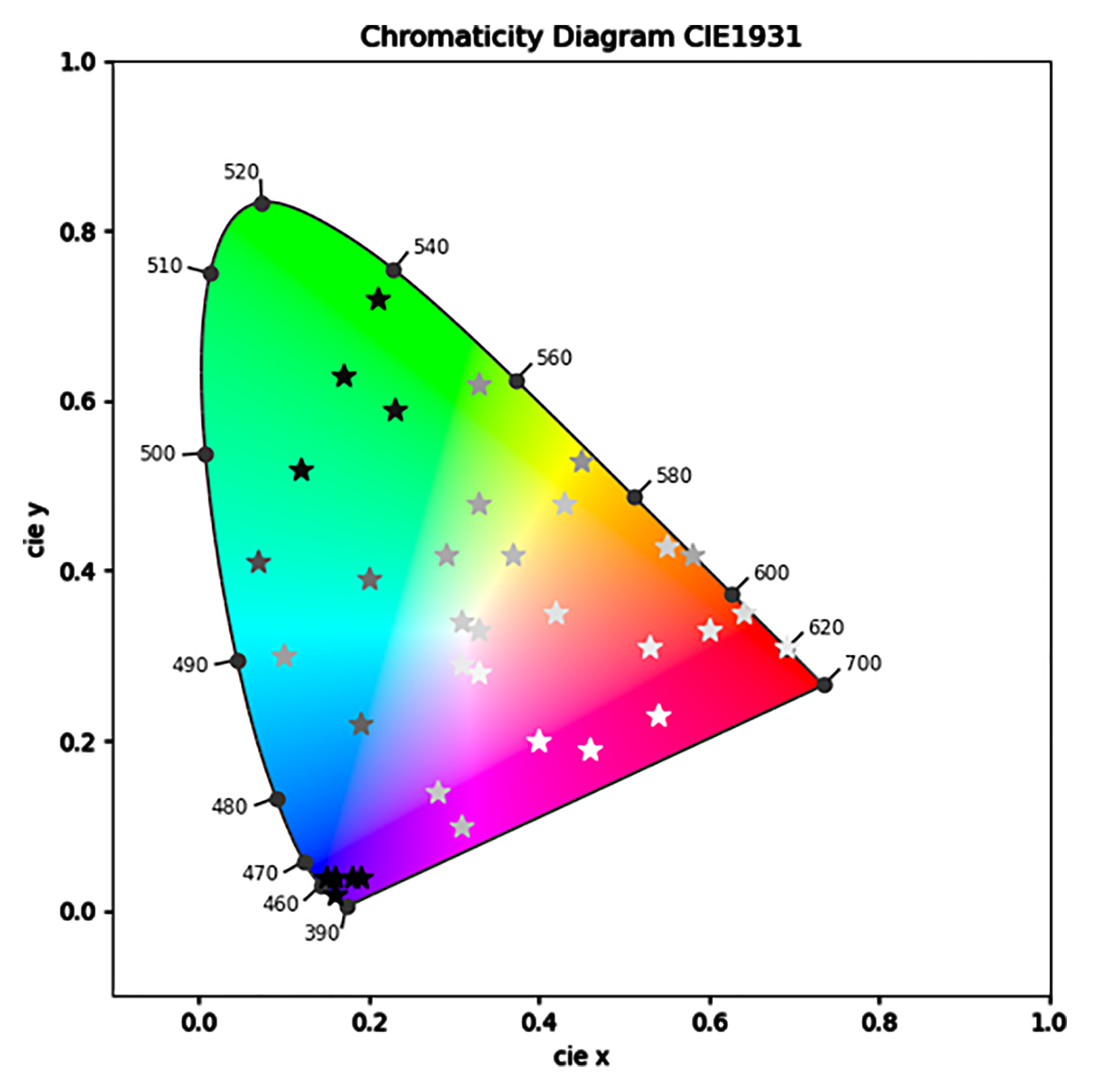}
	}
	\subfloat[All Color Pencil]{
		\includegraphics[width=5.2cm]{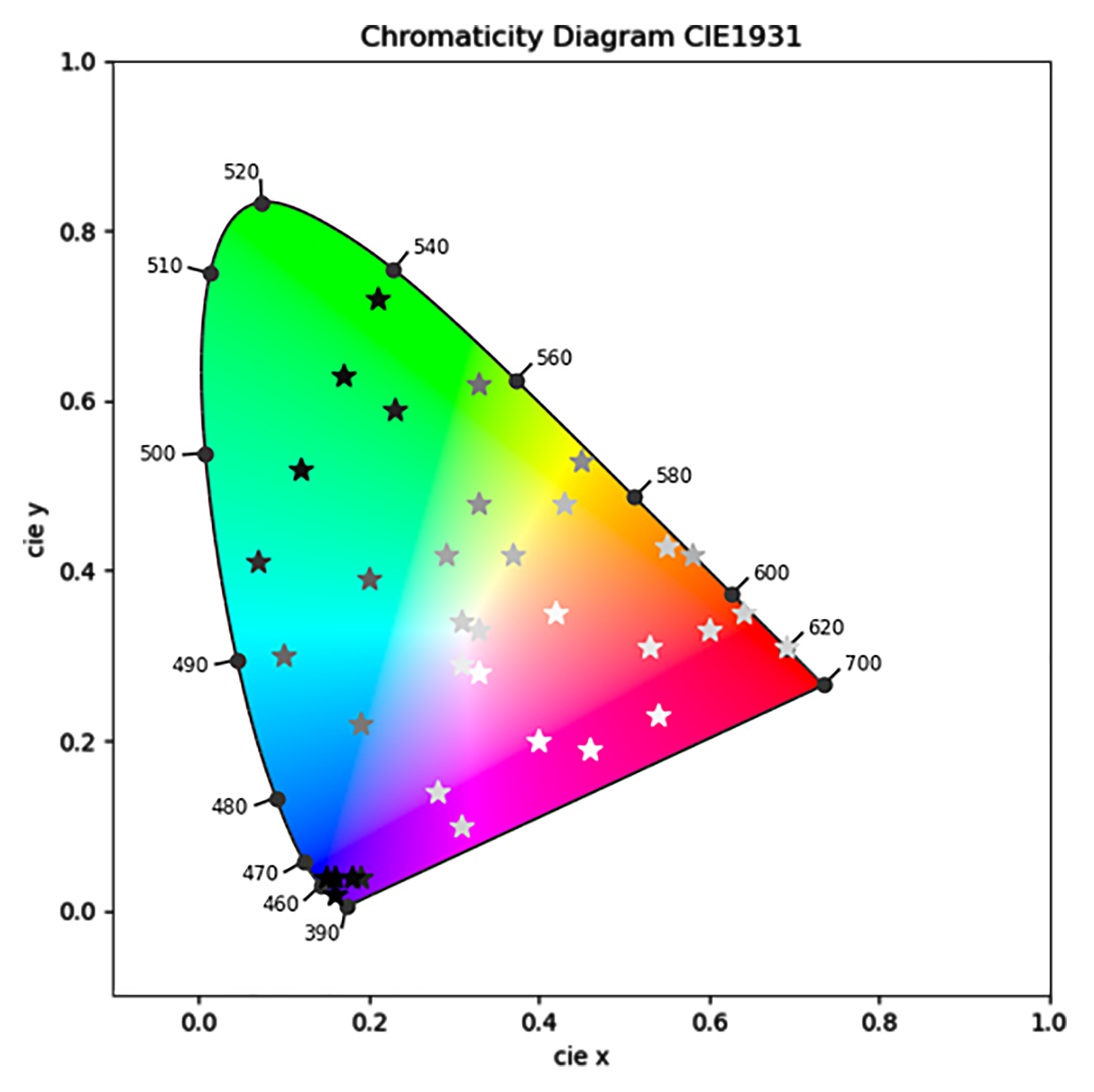}
	}
	\caption{The accuracy on the CIE 1931 Chromaticity Diagram under AlexNet,  where the higher brightness, the higher accuracy. }
	\label{alexnet}
\end{figure}

Part of the accuracy of monochromatic and all color pencils under different multi-spectral light sources are shown in TABLE~\ref{table:alexnet}, which are the top 20 ranked according to the accuracy of all color pencils, and the complete data can be found in Appendix~\ref{appen:alexnet}. One can see that the performance of different multi-spectral light sources on different color pencils is different.  
\begin{table}[htbp]
	\renewcommand\tabcolsep{15.0pt}
	\begin{tabular}{ccccccc}
		\toprule
		\textbf{Color Coordinate} & \textbf{Red} & \textbf{Green} & \textbf{Blue} & \textbf{Yellow} & \textbf{Purple} & \textbf{All Colors} \\
		\midrule
		(0.40, 0.20) & 1.000   & 1.000   & 0.962   & 1.000   & 1.000   & 0.992 \\
		
		(0.33, 0.28) & 1.000   & 1.000   & 1.000   & 1.000   & 0.962   & 0.992 \\
		
		(0.46, 0.19) & 1.000   & 0.923   & 1.000   & 1.000   & 1.000   & 0.985 \\
		
		(0.42, 0.35) & 1.000   & 1.000   & 1.000   & 1.000   & 0.885   & 0.977 \\
		
		(0.54, 0.23) & 1.000   & 0.885   & 0.923   & 1.000   & 1.000   & 0.962 \\
		
		(0.31, 0.29) & 1.000   & 0.769   & 0.962   & 1.000   & 0.923   & 0.931 \\
		
		(0.53, 0.31) & 1.000   & 0.808   & 0.769   & 1.000   & 0.923   & 0.900 \\
		
		(0.33, 0.33) & 1.000   & 0.577   & 0.885   & 1.000   & 0.846   & 0.862 \\
		
		(0.28, 0.14) & 1.000   & 1.000   & 0.462   & 1.000   & 0.769   & 0.846 \\
		
		(0.60, 0.33) & 1.000   & 0.769   & 0.538   & 1.000   & 0.885   & 0.838 \\
		
		(0.69, 0.31) & 1.000   & 0.731   & 0.462   & 1.000   & 0.923   & 0.823 \\
		
		(0.31, 0.10) & 1.000   & 0.962   & 0.385   & 1.000   & 0.731   & 0.815 \\
		
		(0.31, 0.34) & 1.000   & 0.385   & 0.885   & 1.000   & 0.808   & 0.815 \\
		
		(0.64, 0.35) & 1.000   & 0.731   & 0.462   & 1.000   & 0.846   & 0.808 \\
		
		(0.55, 0.43) & 1.000   & 0.731   & 0.462   & 1.000   & 0.808   & 0.800 \\
		
		(0.43, 0.48) & 1.000   & 0.500   & 0.423   & 1.000   & 0.769   & 0.738 \\
		
		(0.37, 0.42) & 1.000   & 0.346   & 0.577   & 1.000   & 0.731   & 0.731 \\
		
		(0.58, 0.42) & 1.000   & 0.462   & 0.346   & 1.000   & 0.654   & 0.692 \\
		
		(0.29, 0.42) & 0.923   & 0.192   & 0.462   & 1.000   & 0.654   & 0.646 \\
		
		(0.33, 0.48) & 1.000   & 0.038   & 0.308   & 0.923   & 0.654   & 0.585 \\
		\bottomrule
	\end{tabular}%
	\caption{Part of the accuracy of monochromatic and all color pencils under different multi-spectral light sources under AlexNet.}
	\label{table:alexnet}%
\end{table}%
The accuracy can also be displayed on the CIE 1931 Chromaticity Diagram, which is shown in FIG.~\ref{alexnet}, where the color of the pentagram represents the accuracy, and the higher brightness, the higher accuracy. 

The point (0.33,0.33) deserves to be noticed, which is the pure white light source. It can be seen that whether for monochromatic or all color pencils, there are always some non-pure white light sources with the accuracy higher than or equal to that of the pure white light. From the view of all color pencils, the accuracy of some non-pure white light sources can increase up to 13\% than that of pure white light. In addition, each monochromatic pencil has a different optimal multi-spectral light source. 

\subsection{VGG19}
\begin{figure}[ht]
	\centering
	\subfloat[Red Pencil]{
		\includegraphics[width=5.2cm]{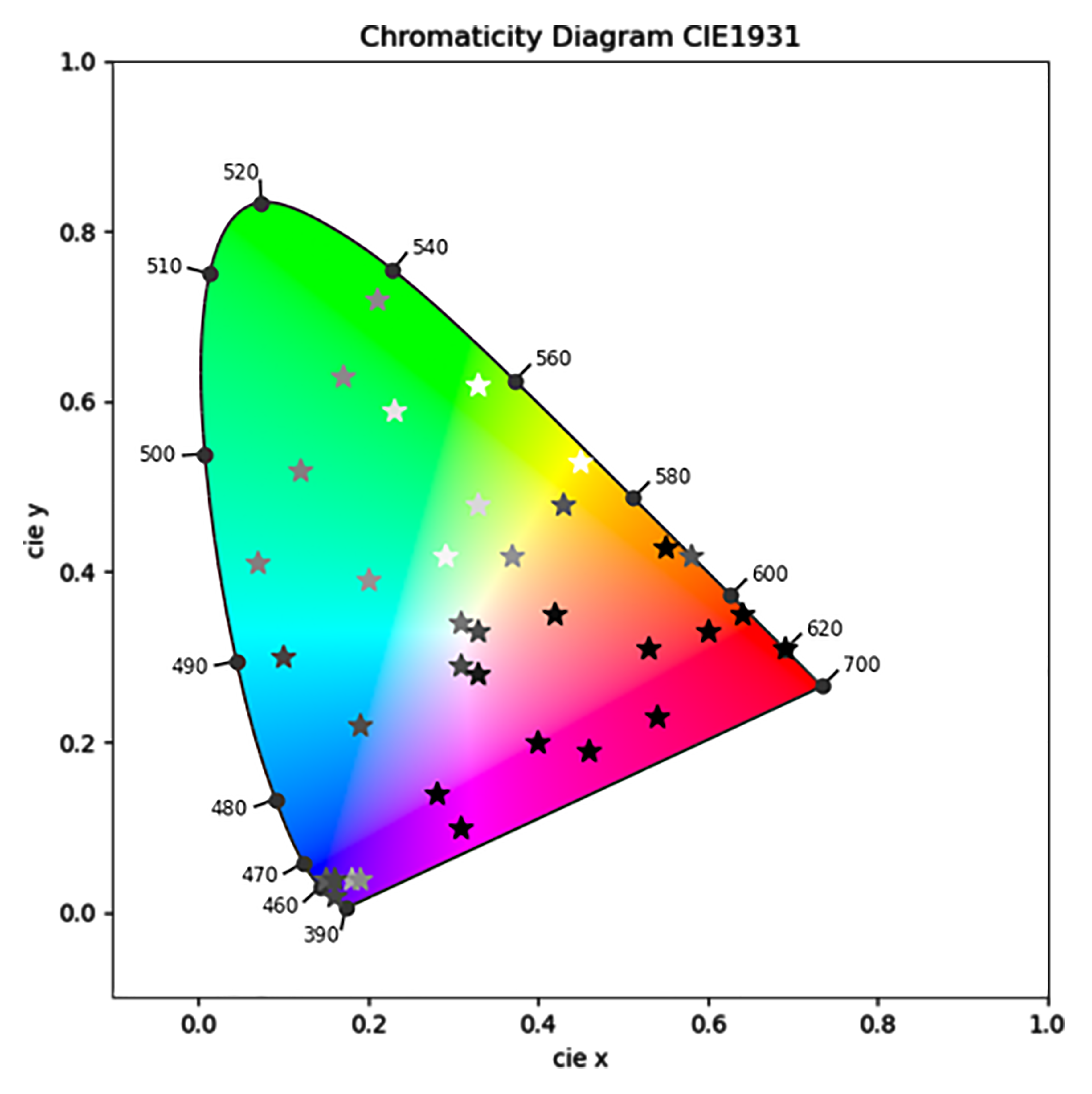}
	}
	\subfloat[Green Pencil]{
		\includegraphics[width=5.2cm]{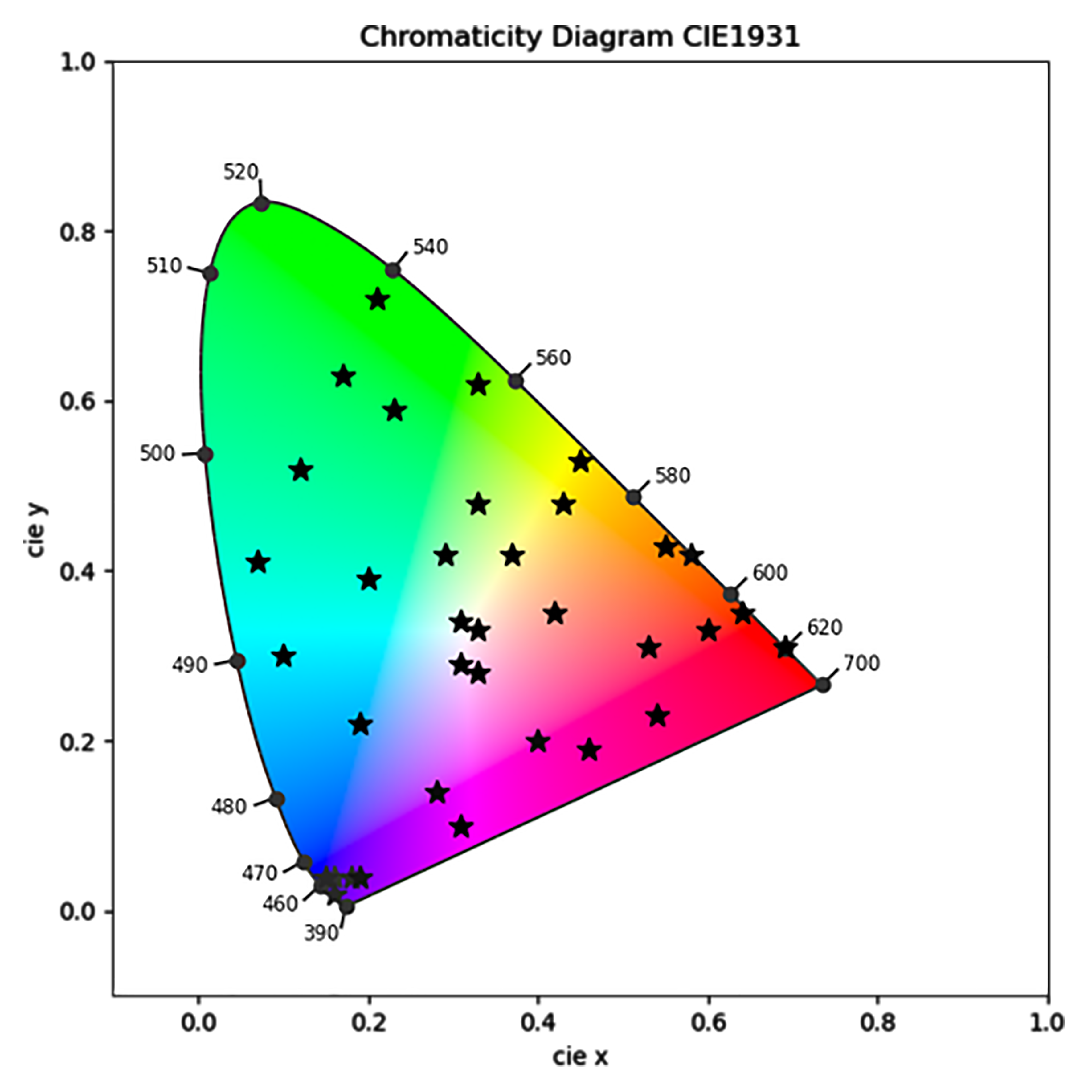}
	}
	\subfloat[Blue Pencil]{
		\includegraphics[width=5.2cm]{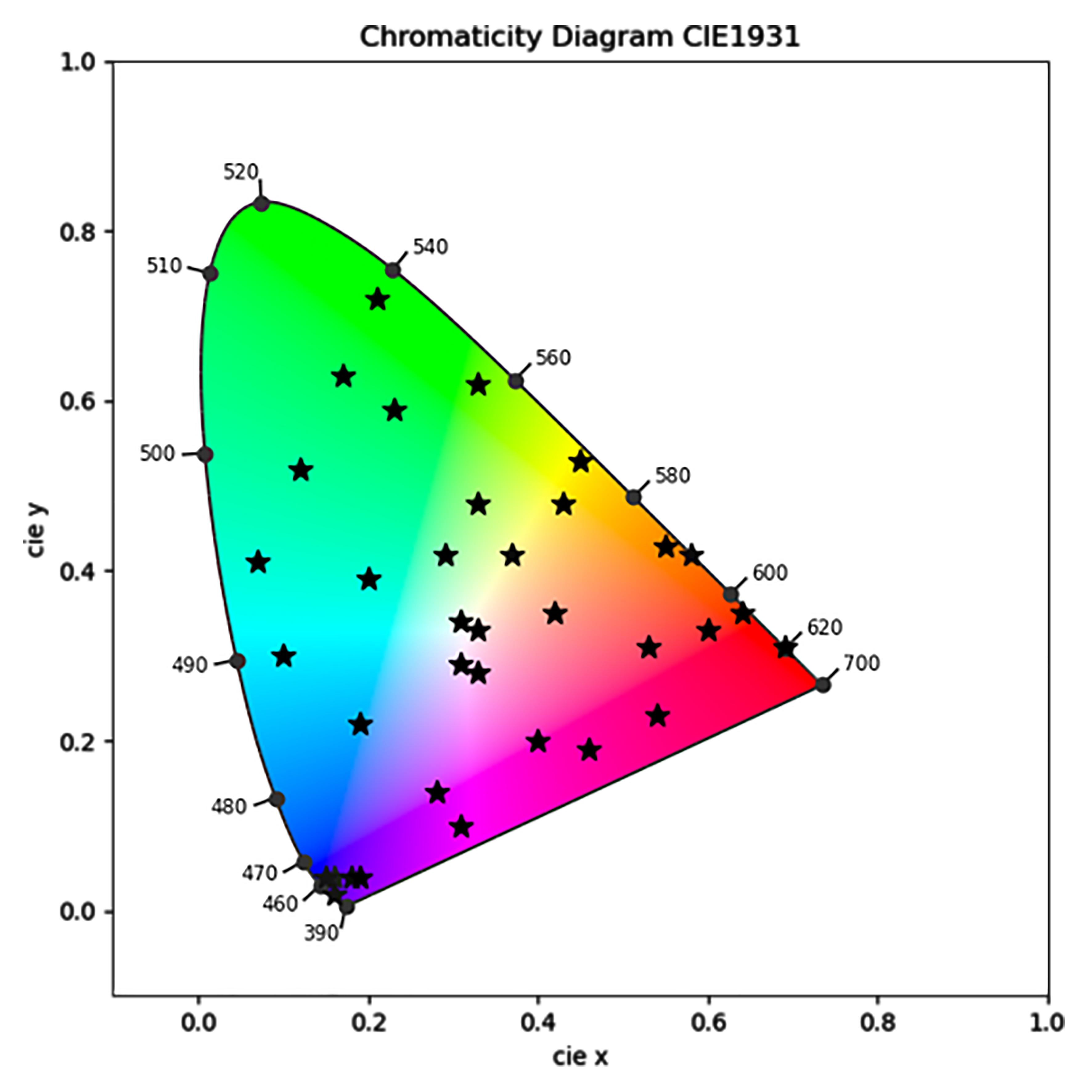}
	}\\
	\subfloat[Yellow Pencil]{
		\includegraphics[width=5.2cm]{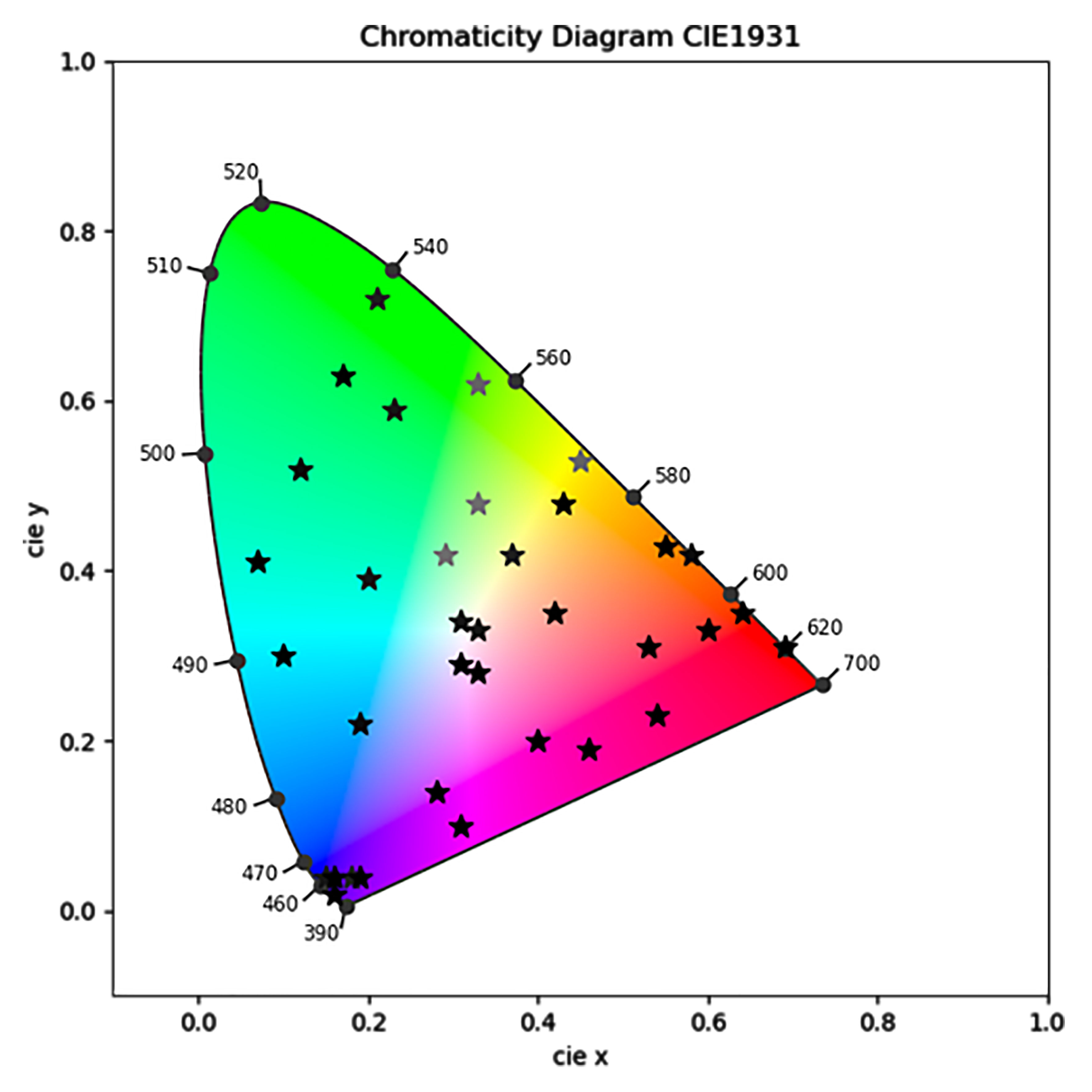}
	}
	\subfloat[Purple Pencil]{
		\includegraphics[width=5.2cm]{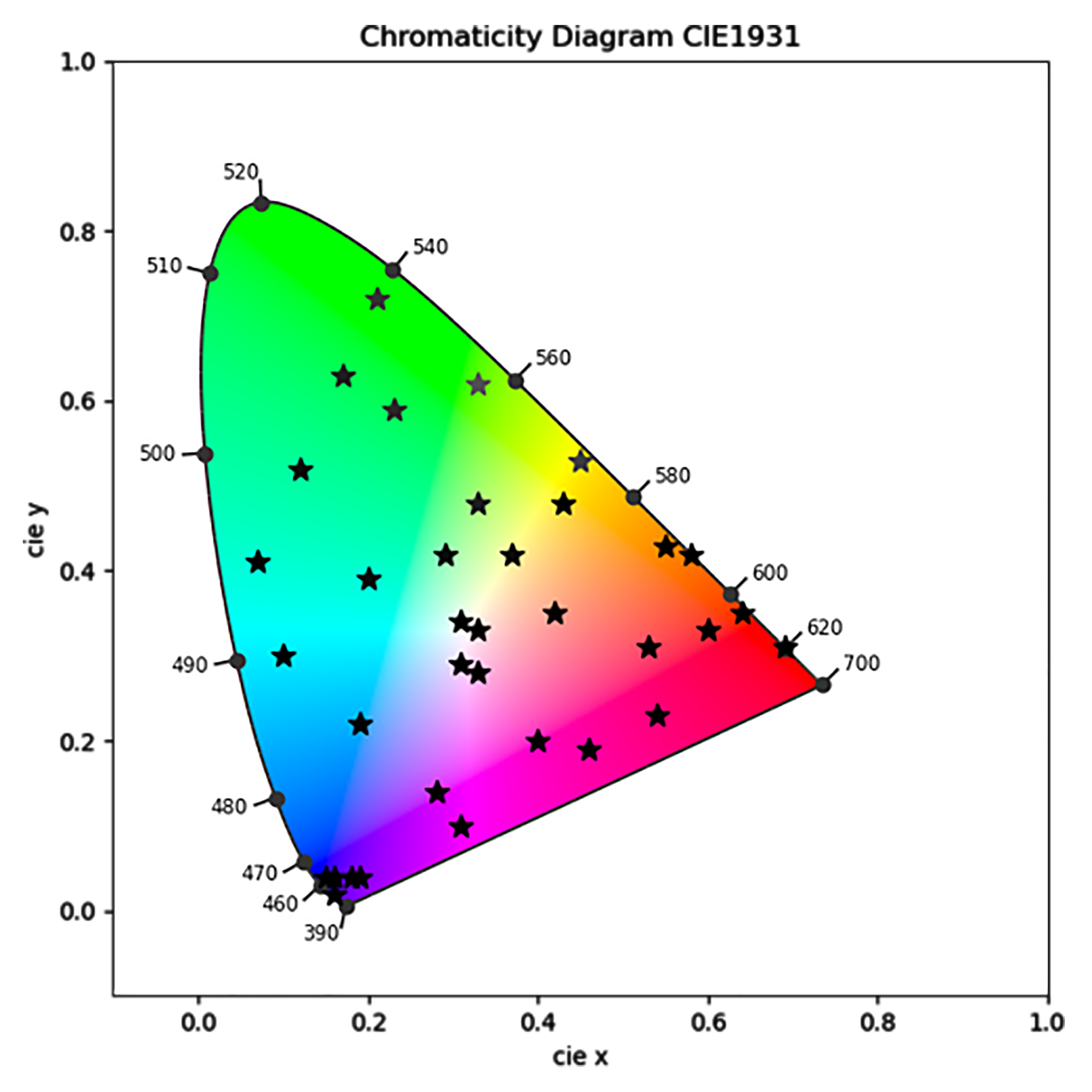}
	}
	\subfloat[All Color Pencil]{
		\includegraphics[width=5.2cm]{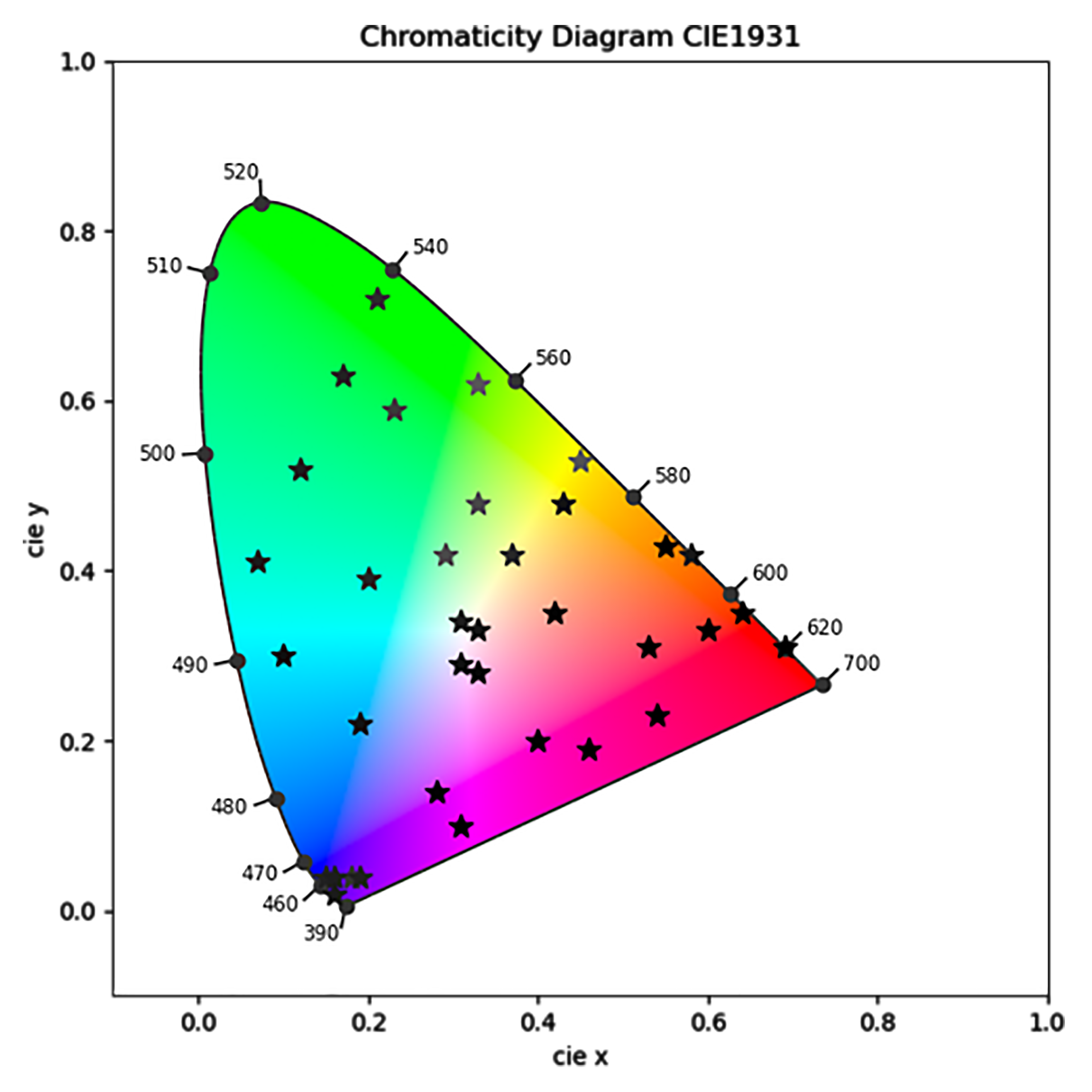}
	}
	\caption{The accuracy on the CIE 1931 Chromaticity Diagram under VGG19,  where the higher brightness, the higher accuracy. }
	\label{vgg19}
\end{figure}

Part of the accuracy is also shown in TABLE~\ref{table:vgg19}, and the complete data can be found in Appendix~\ref{appen:vgg19}. The accuracy on the CIE 1931 Chromaticity Diagram under VGG19 is shown in FIG.~\ref{vgg19}.
\begin{table}[htbp]
	\renewcommand\tabcolsep{15.0pt}
	\begin{tabular}{ccccccc}
		\toprule
		\textbf{Color Coordinate} & \textbf{Red} & \textbf{Green} & \textbf{Blue} & \textbf{Yellow} & \textbf{Purple} & \textbf{All Colors} \\
		\midrule
		(0.33, 0.62) & 1.000   & 0.000   & 0.000   & 0.385 & 0.308 & 0.338 \\
		
		(0.45, 0.53) & 1.000   & 0.000   & 0.000   & 0.385 & 0.231 & 0.323 \\
	
		(0.29, 0.42) & 0.962   & 0.000   & 0.000   & 0.423 & 0.038 & 0.285 \\
		
		(0.33, 0.48) & 0.846   & 0.000   & 0.000   & 0.423 & 0.115 & 0.277 \\
		
		(0.23, 0.59) & 0.885   & 0.000   & 0.000   & 0.077 & 0.154 & 0.223 \\
		
		(0.18, 0.04) & 0.654   & 0.154   & 0.038   & 0.115 & 0.000 & 0.192 \\
		
		(0.21, 0.72) & 0.538   & 0.000   & 0.000   & 0.115 & 0.192 & 0.169 \\
		
		(0.37, 0.42) & 0.577   & 0.000   & 0.038   & 0.115 & 0.038 & 0.154 \\
		
		(0.20, 0.39) & 0.577   & 0.000   & 0.000   & 0.077 & 0.038 & 0.138 \\
		
		(0.15, 0.04) & 0.346   & 0.154   & 0.077   & 0.077 & 0.000 & 0.131 \\
		
		(0.17, 0.63) & 0.538   & 0.000   & 0.000   & 0.000 & 0.115 & 0.131 \\
		
		(0.19, 0.04) & 0.538   & 0.077   & 0.000   & 0.000 & 0.000 & 0.123 \\
		
		(0.12, 0.52) & 0.500   & 0.000   & 0.000   & 0.038 & 0.038 & 0.115 \\
		
		(0.16, 0.04) & 0.269   & 0.192   & 0.115   & 0.000 & 0.000 & 0.115 \\
		
		(0.31, 0.34) & 0.462   & 0.000   & 0.000   & 0.077 & 0.000 & 0.108 \\
		
		(0.07, 0.41) & 0.500   & 0.000   & 0.000   & 0.000 & 0.000 & 0.100 \\
		
		(0.16, 0.02) & 0.269   & 0.154   & 0.077   & 0.000 & 0.000 & 0.100 \\
		
		(0.58, 0.42) & 0.346   & 0.038   & 0.038   & 0.038 & 0.000 & 0.092 \\
		
		(0.33, 0.33) & 0.308   & 0.000   & 0.000   & 0.077 & 0.000 & 0.077 \\
		
		(0.19, 0.22) & 0.308   & 0.000   & 0.000   & 0.038 & 0.000 & 0.069 \\
		\bottomrule
	\end{tabular}%
	\caption{Part of the accuracy of monochromatic and all color pencils under different multi-spectral light sources under VGG19.}
	\label{table:vgg19}%
\end{table}%

The same phenomenon, where the pure white light is non-optimal and the optimal multi-spectral light source of the different monochromatic pencil is different, that occurs in AlexNet also occurs in VGG19. From the view of all color pencils, the accuracy of some non-pure white light sources can increase up to 26\% than that of pure white light. However, the optimal multi-spectra of the two models are different, which means that the optimal multi-spectra are model-dependent. Furthermore, it is surprising that the overall performance of AlexNet is better than that of VGG19. It is accepted that VGG19 should work better than AlexNet, for VGG19 has a larger number of parameters and a more complex network structure, and the performance of VGG19 is better than that of AlexNet on the original ImageNet-1K datasets, in which the top 1 accuracy of AlexNet is 56.5\%, and VGG19 is 74.2\%. This anomaly is left for the following discussion. 

\section{Discussion}\label{sec:discussion}
The multi-spectral light source with the color coordinate of (0.33, 0.33) is the light source we need to focus on because it is pure white light. The comparative effect of non-pure white light sources and pure white light source is the focus of our research and the significance of this study. From the results of the previous two models, we can see that 
whether for monochromatic or all color pencils, there are always some non-pure white light source effects greater than or equal to the pure white light. This suggests that machine vision light sources are not necessarily best in pure white light, and that research on multi-spectral light sources has the potential to further enhance the effectiveness of machine vision and merits further study. This fact is contrary to general common sense, because, for human vision, it is generally the best white light, which again illustrates the difference between machine vision and human vision. 

Meanwhile, this phenomenon is also understandable, because deep neural networks are extracted through convolutional operations of image features, and certain channels may be more important. Some non-pure white light sources can highlight the information of these channels, and the purer information, the more conducive to machine vision image recognition.

The problem of model-dependent of the optimal multi-spectral light source is relatively well understood: different models have different focuses on the utilization of different spectral channels in identification, and therefore the most favorable multi-spectral light source will vary. 

It is noteworthy that the overall performance of VGG19 is lower than that of AlexNet in the multi-spectral light source images. For example, the best result that can be achieved in the recognition of all color pencils, AlexNet is 99.2\%, while VGG19 is 33.8\%. One possible reason is that VGG19 undertrains labels similar to 'pencil', i.e., undertrains training sets with labels 'ballpoint', 'fountain\_pen', 'matchstick', and 'pole', which are themselves poorly recognized. We perform the same recognition operations on these labeled training set images as in this study, and the accuracy is shown in TABLE~\ref{table:imagenet}. 
\begin{table}[htbp]
	\renewcommand\tabcolsep{15.0pt}
	\begin{tabular}{cccccc}
		\toprule
		\textbf{Model} & \textbf{ballpoint} & \textbf{fountain\_pen} & \textbf{matchstick} & \textbf{pole} & \textbf{All Image} \\
		\midrule
		ALexNet & 0.608 & 0.742   & 0.688   & 0.425 &  0.617 \\
		
		VGG19 & 0.794 & 0.946   & 0.878   & 0.521 &  0.786 \\
		\bottomrule
	\end{tabular}%
	\caption{The accuracy of the same recognition operations on the training set images with labels similar to 'pencil'. }
	\label{table:imagenet}%
\end{table}%
As one can see, this reason is negated, and the performance is normal on the training set images with labels similar to 'pencil'. Another possible reason is the effect of illumination of the object surface, the illumination in the experiment is not suitable for the recognition of VGG19, or VGG really does not perform well in such scenes. The peculiarities of pencils may also be a reason to be considered, which requires the study of more objects. 

However, regardless of the reason, it shows the importance of the choice of light sources and models. The right light source and model will greatly enhance the machine vision effect. The reasons for the differences in the effects between models and the criteria for model selection are left for further study. 

\section{Summary}\label{sec:summary}
In summary, the performance of multi-spectral light sources on different object colors in machine vision is studied. Fixing the illuminance value, 35 different multi-spectral light sources are chosen and by recognizing the collected images in these different multi-spectral light sources, the comparison of machine vision effects is done. The results show that whether for monochromatic or all color pencils, there are always some non-pure white light source effects greater than or equal to pure white light. For general color, the accuracy of non-pure white light sources can get improved compared to pure white light, which suggests the potential of multi-spectral light sources to further enhance the effectiveness of machine vision. Two models, AlexNet and VGG19, are compared, and surprised to find that the overall performance of VGG19 is lower than that of AlexNet, which shows the importance of the choice of light sources and models. 

It should be pointed out that although there is some discussion of the causes of these phenomena, the deeper causes will certainly involve the interpretability of neural networks. In addition, the criteria for model selection under multi-spectral light sources are not discussed. Furthermore, this study is an experiment conducted on pencils, and whether the conclusions can be generalized is something that needs to be carefully studied. As a preliminary study of the role of multi-spectral light sources in machine vision, these problems need to be discussed in more detail.

\begin{acknowledgments}
	
This work was supported by Beijing Postdoctoral Research Foundation.
	
\end{acknowledgments}

\clearpage

\appendix
\section{The complete accuracy of different multi-spectral light sources under AlexNet \label{appen:alexnet}}
\begin{table}[htbp]
	\renewcommand\tabcolsep{15.0pt}
	\begin{tabular}{ccccccc}
		\toprule
		\textbf{Color Coordinate} & \textbf{Red} & \textbf{Green} & \textbf{Blue} & \textbf{Yellow} & \textbf{Purple} & \textbf{All Colors} \\
		\midrule
		(0.40, 0.20) & 1.000   & 1.000   & 0.962   & 1.000   & 1.000   & 0.992 \\
		
		(0.33, 0.28) & 1.000   & 1.000   & 1.000   & 1.000   & 0.962   & 0.992 \\
		
		(0.46, 0.19) & 1.000   & 0.923   & 1.000   & 1.000   & 1.000   & 0.985 \\
		
		(0.42, 0.35) & 1.000   & 1.000   & 1.000   & 1.000   & 0.885   & 0.977 \\
		
		(0.54, 0.23) & 1.000   & 0.885   & 0.923   & 1.000   & 1.000   & 0.962 \\
		
		(0.31, 0.29) & 1.000   & 0.769   & 0.962   & 1.000   & 0.923   & 0.931 \\
		
		(0.53, 0.31) & 1.000   & 0.808   & 0.769   & 1.000   & 0.923   & 0.900 \\
		
		(0.33, 0.33) & 1.000   & 0.577   & 0.885   & 1.000   & 0.846   & 0.862 \\
		
		(0.28, 0.14) & 1.000   & 1.000   & 0.462   & 1.000   & 0.769   & 0.846 \\
		
		(0.60, 0.33) & 1.000   & 0.769   & 0.538   & 1.000   & 0.885   & 0.838 \\
		
		(0.69, 0.31) & 1.000   & 0.731   & 0.462   & 1.000   & 0.923   & 0.823 \\
		
		(0.31, 0.10) & 1.000   & 0.962   & 0.385   & 1.000   & 0.731   & 0.815 \\
		
		(0.31, 0.34) & 1.000   & 0.385   & 0.885   & 1.000   & 0.808   & 0.815 \\
	
		(0.64, 0.35) & 1.000   & 0.731   & 0.462   & 1.000   & 0.846   & 0.808 \\
		
		(0.55, 0.43) & 1.000   & 0.731   & 0.462   & 1.000   & 0.808   & 0.800 \\
		
		(0.43, 0.48) & 1.000   & 0.500   & 0.423   & 1.000   & 0.769   & 0.738 \\
		
		(0.37, 0.42) & 1.000   & 0.346   & 0.577   & 1.000   & 0.731   & 0.731 \\
		
		(0.58, 0.42) & 1.000   & 0.462   & 0.346   & 1.000   & 0.654   & 0.692 \\
		
		(0.29, 0.42) & 0.923   & 0.192   & 0.462   & 1.000   & 0.654   & 0.646 \\
		
		(0.33, 0.48) & 1.000   & 0.038   & 0.308   & 0.923   & 0.654   & 0.585 \\
		
		(0.45, 0.53) & 0.962   & 0.038   & 0.192   & 1.000   & 0.577   & 0.554 \\
		
		(0.19, 0.22) & 1.000   & 0.115   & 0.423   & 0.423   & 0.385   & 0.469 \\
		
		(0.33, 0.62) & 1.000   & 0.000   & 0.000   & 0.654   & 0.577   & 0.446 \\
		
		(0.10, 0.30) & 0.615   & 0.077   & 0.615   & 0.115   & 0.615   & 0.408 \\
		
		(0.20, 0.39) & 0.885   & 0.000   & 0.500   & 0.038   & 0.423   & 0.369 \\
		
		(0.07, 0.41) & 0.385   & 0.000   & 0.308   & 0.000   & 0.308   & 0.200 \\
		
		(0.19, 0.04) & 0.346   & 0.000   & 0.000   & 0.346   & 0.000   & 0.138 \\
		
		(0.23, 0.59) & 0.577   & 0.000   & 0.000   & 0.000   & 0.077   & 0.131 \\
		
		(0.12, 0.52) & 0.269   & 0.000   & 0.000   & 0.000   & 0.038   & 0.062 \\
		
		(0.17, 0.63) & 0.231   & 0.000   & 0.000   & 0.000   & 0.038   & 0.054 \\
		
		(0.21, 0.72) & 0.231   & 0.000   & 0.000   & 0.000   & 0.038   & 0.054 \\
		
		(0.18, 0.04) & 0.038   & 0.000   & 0.000   & 0.077   & 0.000   & 0.023 \\
		
		(0.15, 0.04) & 0.000   & 0.000   & 0.000   & 0.000   & 0.000   & 0.000 \\
		
		(0.16, 0.02) & 0.000   & 0.000   & 0.000   & 0.000   & 0.000   & 0.000 \\
		
		(0.16, 0.04) & 0.000   & 0.000   & 0.000   & 0.000   & 0.000   & 0.000 \\
		\bottomrule
	\end{tabular}%
\end{table}%
\clearpage
\section{The complete accuracy of different multi-spectral light sources under VGG19 \label{appen:vgg19}}
\begin{table}[htbp]
	\renewcommand\tabcolsep{15.0pt}
	\begin{tabular}{ccccccc}
		\toprule
		\textbf{Color Coordinate} & \textbf{Red} & \textbf{Green} & \textbf{Blue} & \textbf{Yellow} & \textbf{Purple} & \textbf{All Colors} \\
		\midrule
		(0.33, 0.62) & 1.000   & 0.000   & 0.000   & 0.385 & 0.308 & 0.338 \\
	
		(0.45, 0.53) & 1.000   & 0.000   & 0.000   & 0.385 & 0.231 & 0.323 \\
		
		(0.29, 0.42) & 0.962   & 0.000   & 0.000   & 0.423 & 0.038 & 0.285 \\
		
		(0.33, 0.48) & 0.846   & 0.000   & 0.000   & 0.423 & 0.115 & 0.277 \\
		
		(0.23, 0.59) & 0.885   & 0.000   & 0.000   & 0.077 & 0.154 & 0.223 \\
		
		(0.18, 0.04) & 0.654   & 0.154   & 0.038   & 0.115 & 0.000 & 0.192 \\
		
		(0.21, 0.72) & 0.538   & 0.000   & 0.000   & 0.115 & 0.192 & 0.169 \\
		
		(0.37, 0.42) & 0.577   & 0.000   & 0.038   & 0.115 & 0.038 & 0.154 \\
		
		(0.20, 0.39) & 0.577   & 0.000   & 0.000   & 0.077 & 0.038 & 0.138 \\
		
		(0.15, 0.04) & 0.346   & 0.154   & 0.077   & 0.077 & 0.000 & 0.131 \\
		
		(0.17, 0.63) & 0.538   & 0.000   & 0.000   & 0.000 & 0.115 & 0.131 \\
		
		(0.19, 0.04) & 0.538   & 0.077   & 0.000   & 0.000 & 0.000 & 0.123 \\
		
		(0.12, 0.52) & 0.500   & 0.000   & 0.000   & 0.038 & 0.038 & 0.115 \\
		
		(0.16, 0.04) & 0.269   & 0.192   & 0.115   & 0.000 & 0.000 & 0.115 \\
		
		(0.31, 0.34) & 0.462   & 0.000   & 0.000   & 0.077 & 0.000 & 0.108 \\
		
		(0.07, 0.41) & 0.500   & 0.000   & 0.000   & 0.000 & 0.000 & 0.100 \\
		
		(0.16, 0.02) & 0.269   & 0.154   & 0.077   & 0.000 & 0.000 & 0.100 \\
		
		(0.58, 0.42) & 0.346   & 0.038   & 0.038   & 0.038 & 0.000 & 0.092 \\
		
		(0.33, 0.33) & 0.308   & 0.000   & 0.000   & 0.077 & 0.000 & 0.077 \\
		
		(0.19, 0.22) & 0.308   & 0.000   & 0.000   & 0.038 & 0.000 & 0.069 \\
		
		(0.43, 0.48) & 0.346   & 0.000   & 0.000   & 0.000 & 0.000 & 0.069 \\
		
		(0.31, 0.29) & 0.308   & 0.000   & 0.000   & 0.000 & 0.000   & 0.062 \\
		
		(0.10, 0.30) & 0.269   & 0.000   & 0.000   & 0.000 & 0.000   & 0.054 \\
		
		(0.33, 0.28) & 0.115   & 0.000   & 0.000   & 0.000 & 0.000   & 0.023 \\
		
		(0.40, 0.20) & 0.000   & 0.038   & 0.000   & 0.000 & 0.000   & 0.008 \\
		
		(0.42, 0.35) & 0.038   & 0.000   & 0.000   & 0.000 & 0.000   & 0.008 \\
		
		(0.28, 0.14) & 0.000   & 0.038   & 0.000   & 0.000 & 0.000   & 0.008 \\
		
		(0.53, 0.31) & 0.000   & 0.000   & 0.000   & 0.000 & 0.000   & 0.000 \\
		
		(0.60, 0.33) & 0.000   & 0.000   & 0.000   & 0.000 & 0.000   & 0.000 \\
		
		(0.31, 0.10) & 0.000   & 0.000   & 0.000   & 0.000 & 0.000   & 0.000 \\
		
		(0.69, 0.31) & 0.000   & 0.000   & 0.000   & 0.000 & 0.000   & 0.000 \\
		
		(0.46, 0.19) & 0.000   & 0.000   & 0.000   & 0.000 & 0.000   & 0.000 \\
		
		(0.54, 0.23) & 0.000   & 0.000   & 0.000   & 0.000 & 0.000   & 0.000 \\
		
		(0.64, 0.35) & 0.000   & 0.000   & 0.000   & 0.000 & 0.000   & 0.000 \\
		
		(0.55, 0.43) & 0.000   & 0.000   & 0.000   & 0.000 & 0.000   & 0.000 \\
		\bottomrule
	\end{tabular}%
	\label{tab:addlabel}%
\end{table}%

\clearpage

\bibliography{refs.bib}
\end{document}